# Constraining Cosmology with Double-source-plane Strong Gravitational Lenses from the AGEL Survey

Duncan J. Bowden[1,2], Nandini Sahu[1,3,4], Anowar J. Shajib[5,6,7,21], Kim-Vy Tran[1,4], Tania M. Barone[4,8], Keerthi Vasan G. C.[9], Daniel J. Ballard[10], Thomas E. Collett[11], Faith Dalessandro[12], Giovanni Ferrami[13], Karl Glazebrook[4,8], William J. Gottemoller[1,14], Leena Iwamoto[1,14], Tucker Jones[12], Glenn G. Kacprzak[4,8], Geraint F. Lewis[10], Haven McIntosh-Lombardo[1,15], Hannah Skobe[5,16], Sherry H. Suyu[17,18], and Sarah M. Sweet[4,19]

[1] Center for Astrophysics, Harvard & Smithsonian, Cambridge, MA 02138, USA; duncan.bowden@cfa.harvard.edu
[2] School of Physics and Astronomy, University of Southampton, Southampton, SO17 1BJ, UK
[3] University of New South Wales, Sydney, NSW, 2033, Australia
[4] ARC center of Excellence for All Sky Astrophysics in 3 Dimensions (ASTRO 3D), Australia
[5] Department of Astronomy & Astrophysics, University of Chicago, Chicago, IL 60637, USA
[6] Kavli Institute for Cosmological Physics, University of Chicago, Chicago, IL 60637, USA
[7] Center for Astronomy, Space Science and Astrophysics, Independent University, Bangladesh, Dhaka 1229, Bangladesh
[8] Center for Astrophysics and Supercomputing, Swinburne University of Technology, Hawthorn, Victoria 3122, Australia
[9] The Observatories of the Carnegie Institution for Science, 813 Santa Barbara St., Pasadena, CA 91101, USA
[10] Sydney Institute for Astronomy, School of Physics, A28, The University of Sydney, NSW 2006, Australia
[11] Institute of Cosmology and Gravitation, University of Portsmouth, Burnaby Rd., Portsmouth PO1 3FX, UK
[12] Department of Physics and Astronomy, University of California, Davis, 1 Shields Ave., Davis, CA 95616, USA
[13] School of Physics, University of Melbourne, Parkville, VIC 3010, Australia
[14] Department of Astronomy, Harvard College, Cambridge, MA 02138, USA
[15] Department of Physics, Northeastern University, Boston, MA 02115, USA
[16] McWilliams Center for Cosmology and Astrophysics, Physics Department, Carnegie Mellon University, Pittsburgh, PA 15213, USA
[17] Technical University of Munich, TUM School of Natural Sciences, Physics Department, James-Franck-Straße 1, 85748 Garching, Germany
[18] Max-Planck-Institut für Astrophysik, Karl-Schwarzschild Straße 1, 85748 Garching, Germany
[19] School of Mathematics and Physics, University of Queensland, Brisbane, QLD 4072, Australia

## Abstract

Double-source-plane strong gravitational lenses (DSPLs), with two sources at different redshifts, are independent cosmological probes of the dark energy equation of state parameter $w$ and the matter density parameter $\Omega_{\rm m}$. We present the lens model for the DSPL AGEL035346−170639 and infer cosmological constraints from this system for flat Λ cold dark matter and flat $w$CDM cosmologies. From the joint posterior of $w$ and $\Omega_{\rm m}$ in the flat $w$CDM cosmology, we extract the following median values and 1$\sigma$ uncertainties: $w = -1.52^{+0.49}_{-0.33}$ and $\Omega_{\rm m} = 0.192^{+0.305}_{-0.131}$ from AGEL0353 alone. Combining our measurements with two previously analyzed DSPLs, we present the joint constraint on these parameters from a sample of three, the largest galaxy-scale DSPL sample used for cosmological measurement to date. The combined precision of $w$ from three DSPLs is higher by 15% over AGEL0353 alone. Combining DSPL and cosmic microwave background (CMB) measurements improves the precision of $w$ from CMB-only constraints by 39%, demonstrating the complementarity of DSPLs with the CMB. Despite their promising constraining power, DSPLs are limited by sample size, with only a handful discovered so far. Although ongoing and near-future wide-area sky surveys will increase the number of known DSPLs by up to two orders of magnitude, these systems will still require dedicated high-resolution imaging and spectroscopic follow-ups like those presented in this paper. Our ASTRO 3D Galaxy Evolution with Lenses collaboration is undertaking such follow-up campaigns for several newly discovered DSPLs and will provide cosmological measurements from larger samples of DSPLs in the future.

## 1. Introduction

The simplest model in concordance with observations of the Universe across many scales is the Λ cold dark matter (ΛCDM) cosmology, considered "the standard model of cosmology." The combination of cosmic microwave background (CMB), Type Ia supernovae (SNe), baryon acoustic oscillation (BAO), weak lensing, and galaxy clustering measurements indicates a cosmology in agreement with the ΛCDM model (D. M. Scolnic et al. 2018; Planck Collaboration 2020; DES Collaboration 2022; DESI Collaboration 2025a). However, the ΛCDM model faces many challenges. Some of the most well-known include the lack of direct detection of dark matter (L. Roszkowski et al. 2018), the evidence for the evolution of the dark energy equation of state parameter (DESI Collaboration 2025b), and the Hubble tension in which there is a discrepancy between local and high-redshift measurements of the Hubble constant $H_0$ (E. Di Valentino et al. 2021). These discrepancies and other challenges (e.g., L. Perivolaropoulos & F. Skara 2022) suggest that alternative cosmologies to ΛCDM may be necessary to describe the evolution of the Universe more accurately. We must endeavor to find descriptions of cosmology that address these issues while retaining the agreement of ΛCDM across many measures. To do so, we must improve our measurements of cosmological parameters using new independent probes to differentiate between ΛCDM and alternative models.

[21] NFHP Einstein Fellow.

Strong gravitational lensing is one phenomenon that could solve this problem. Double-source-plane lenses (DSPLs), with source galaxies at two distinct redshifts, are a unique class of strong lenses with a direct dependence on cosmological parameters. Due to the sensitivity of DSPLs to the spatial geometry of the Universe, they are proven probes of cosmology (T. E. Collett & M. W. Auger 2014; R. J. Smith & T. E. Collett 2021; N. Sahu et al. 2025). The ratio of Einstein radii within a DSPL is related to the angular diameter distances up to and between the deflector and source galaxy redshifts (T. E. Collett et al. 2012). Single-source systems require additional data beyond imaging to be usable as cosmological probes, for example, stellar kinematics (e.g., T. Li et al. 2024) and time delays (albeit limited to lenses with variable point sources; e.g., S. Birrer et al. 2025). For the DSPLs, however, the ratio of the Einstein radii can be constrained solely from imaging data. Since angular diameter distances are functions of the matter density parameter $\Omega_{\rm m}$ for a flat ΛCDM cosmology, and the dark energy equation of state parameter $w$ for a more flexible flat $w$CDM cosmology, constraining the ratio of distances allows us to infer $\Omega_{\rm m}$ and $w$ (see Section 2.1).

E. Jullo et al. (2010), G. Caminha et al. (2022) (and others) have shown that it is possible to use multisource-plane galaxy cluster strong lenses to constrain $\Omega_{\rm m}$ and $w$. The many arcs provided by cluster lenses help to break the degeneracy between different mass profiles. However, these systems require deep, high-resolution imaging to identify many sources and spectroscopic follow-ups to measure the redshift of each source, and have complex mass distributions comprised of many galactic and dark matter components (e.g., G. Caminha et al. 2019; G. Mahler et al. 2023). These caveats motivate the use of galaxy-scale multisource-plane strong lenses, which have simple mass distributions (S. Vegetti et al. 2014). T. E. Collett & M. W. Auger (2014) were the first to show that galaxy-scale DSPLs were viable to constrain $\Omega_{\rm m}$ and $w$, using the DSPL SDSSJ0946+1006 (hereafter J0946). However, there are persistent caveats to using these systems, not limited to mass model degeneracies with line-of-sight perturbers (D. Johnson et al. 2025), influences from the local deflector environment, and unknown systematic biases that we hope to address in future work when building a larger sample of DSPLs.

Identifying DSPLs is the first step in the process; however, they must be confirmed with accurate redshift measurements of each component within the system. Spectroscopic surveys excel at this, as exemplified by the ASTRO 3D Galaxy Evolution with Lenses (AGEL) survey (K.-V. H. Tran et al. 2022; T. M. Barone et al. 2025). AGEL's primary goal is to collate a high-quality catalog of strong gravitational lenses with spectroscopic redshift measurements and high-resolution space-based imaging. Redshift measurements validate the presence of objects at multiple redshifts, namely the deflector and the source galaxies. High-resolution imaging further confirms strong lensing candidates by resolving multiple images of the same background source. To mitigate systematic biases, we will select only the highest-quality DSPL observations using integral field spectroscopy, in conjunction with AGEL's ongoing work to characterize the local environment of strong lensing deflectors (W. J. Gottemoller et al. 2025, in preparation).

Historically, DSPLs are rare; only $\mathcal{O}(10)$ have been discovered to date. We are limited to a handful of DSPLs per $10^3$ single galaxy strong lenses observable by upcoming wide-area surveys, such as the Euclid Wide Survey (G. Ferrami & S. Wyithe 2024). However, this limitation is offset by the large number of galaxy-galaxy strong lenses that ongoing and future surveys from Euclid and the Vera Rubin Observatory are expected to observe, which is estimated to be $\mathcal{O}(10^5)$ (T. E. Collett 2015; G. Ferrami & S. Wyithe 2024; A. J. Shajib et al. 2024).

From Euclid's Quick Data Release 1 (Euclid Collaboration 2025a), four DSPL candidates have already been identified across 63.1 deg$^2$ of the sky by the Euclid Collaboration (2025b). These authors predict that ∼1700 DSPL candidates will be detected over the whole Euclid survey, which will cover 14,000 deg$^2$. The 4MOST Strong Lens Spectroscopic Legacy Survey (4SLSLS; T. E. Collett et al. 2023) will obtain spectroscopic redshifts to confirm ∼300 of these DSPL candidates, an order of magnitude greater than the number of DSPLs required to provide uncertainties in $w$ less than 10% (T. E. Collett et al. 2012). Given this large sample of DSPLs, we can also select systems with favorable configurations that lend themselves to cosmography (T. E. Collett et al. 2012), improving our prospects of reducing the total uncertainty. However, we can achieve the same aim by increasing the sample of DSPLs used for cosmography (e.g., D. Sharma et al. 2023; A. J. Shajib et al. 2024), not necessarily only those with favorable source positions. Accordingly, in this paper, we introduce an additional DSPL to the existing sample of two galaxy-scale DSPLs (J0946, AGEL1507) that have been analyzed for cosmographic measurements (T. E. Collett & M. W. Auger 2014; N. Sahu et al. 2025).

Here, we present our lens modeling results for the DSPL AGEL035346−170639 (hereafter AGEL0353) and its corresponding cosmological constraints on $\Omega_{\rm m}$ and $w$ for flat ΛCDM and flat $w$CDM models. In Section 2, we summarize gravitational lensing theory and the dependence of multiplane lensing upon cosmology. In Section 3, we introduce AGEL0353 and describe the imaging and spectroscopic data used in this work. We detail the lens modeling process and present the modeling results in Sections 4 and 5. In Section 6, we present the constraints on $\Omega_{\rm m}$ and $w$ from our model for AGEL0353, and the combined constraints from the three modeled DSPLs, CMB measurements, and other probes such as BAO and SNe. We discuss our results and future prospects in Section 7 and conclusions in Section 8.

## 2. Compound Gravitational Lensing

Strong lensing describes the regime in which the gravitational deflection of light rays from a source results in multiple images. Compound or multiplane lenses are systems with two or more sources at different redshifts, where the intermediate sources also act as deflectors. For compound lenses, a generalized lens equation is required to describe the path light takes from each source. We follow the formalism of

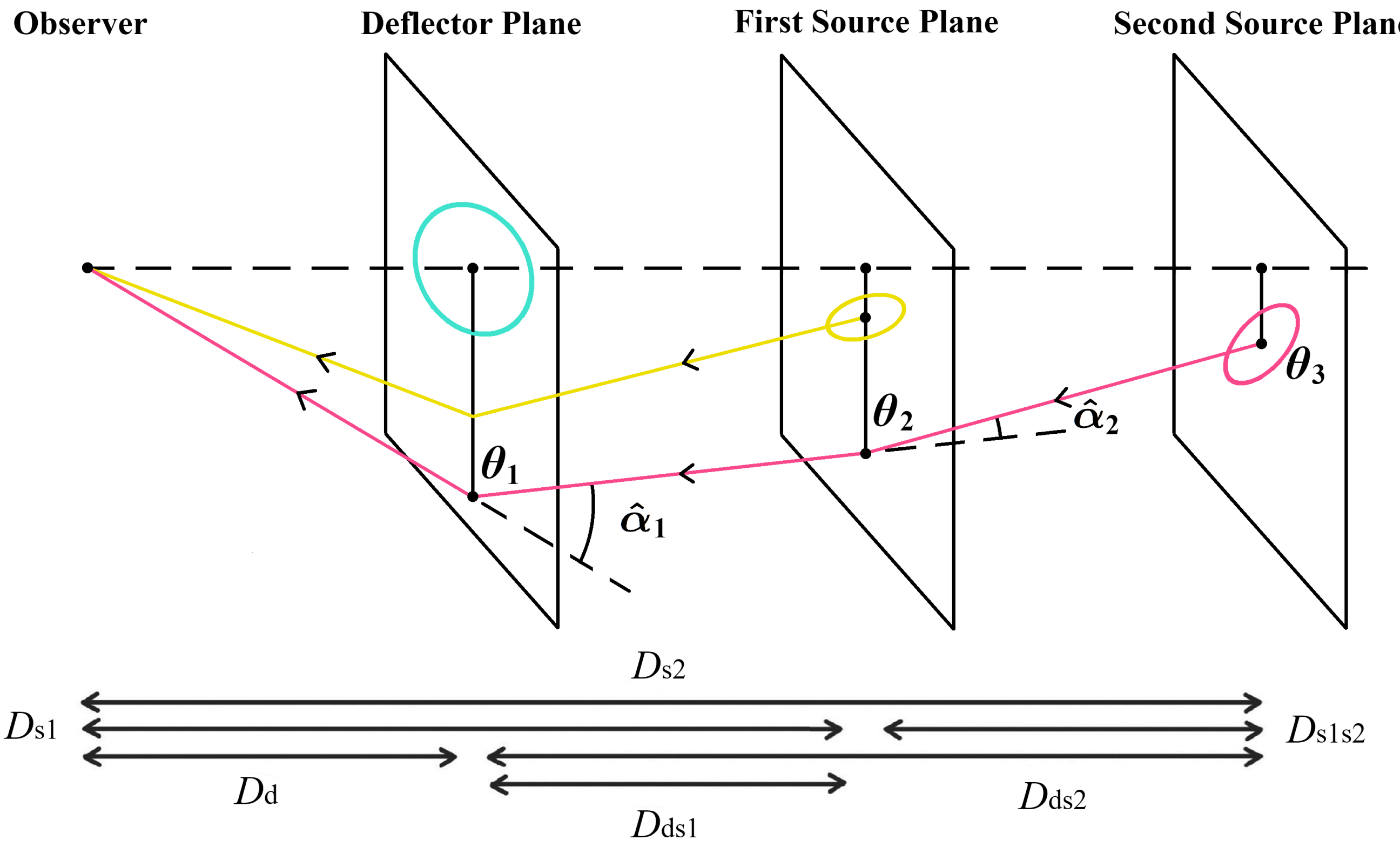

**Figure 1.** Schematic diagram of a double-source-plane lens (DSPL) based on Figure 12 of P. Schneider et al. (2006). Yellow rays show the path of light from source 1, deflected by the deflector toward the observer. Pink rays show the path of light from source 2, deflected by source 1, and the deflector toward the observer. The dashed line represents the line of sight through the center of the deflector. Angular diameter distances $D$ are displayed at the bottom by double-headed arrows, where the subscripts d, s1, and s2 denote the deflector, source 1, and source 2. $\boldsymbol{\theta_1}$, $\boldsymbol{\theta_2}$, and $\boldsymbol{\theta_3}$ denote the angular positions of source 2 in the deflector plane, in the first source plane, and true position in the second source plane. The angles $\hat{\boldsymbol{\alpha}}_\mathbf{1}$ and $\hat{\boldsymbol{\alpha}}_\mathbf{2}$ represent deflection by the deflector and source 1.

P. Schneider et al. (1992) for the case of double-source-plane systems, depicted in Figure 1, with a deflector, (nearer) source 1, and (farther) source 2.

For DSPLs, the lens equation can be written as:

$$\boldsymbol{\theta_3} = \boldsymbol{\theta_1} - \boldsymbol{\alpha}_\mathbf{1}^{\prime}(\boldsymbol{\theta_1}) - \boldsymbol{\alpha}_\mathbf{2}^{\prime}(\boldsymbol{\theta_2}), \tag{1}$$

where $\boldsymbol{\theta_3}$ is the true angular position of source 2, $\boldsymbol{\theta_2} = \boldsymbol{\theta_1} - \beta_{12}\boldsymbol{\alpha}_\mathbf{1}^{\prime}(\boldsymbol{\theta_1})$ represents the angular position in the intermediate plane, and $\boldsymbol{\theta_1}$ is the angular image position observed in the deflector plane. Furthermore, $\boldsymbol{\alpha}_{\boldsymbol{i}}^{\prime}(\boldsymbol{\theta})$ is the scaled deflection angle that is related to the physical deflection angle via $\boldsymbol{\alpha'}_{\boldsymbol{i}}(\boldsymbol{\theta}) = \frac{D_{i3}}{D_3}\hat{\boldsymbol{\alpha}}_{\boldsymbol{i}}(\boldsymbol{\theta})$, with subscripts 1, 2, and 3 denoting the main deflector, source 1, and source 2 planes.

Here, $\beta_{12}$ is the cosmological distance ratio and is defined as:

$$\beta_{12} = \frac{D_{\mathrm{ds1}}\, D_{\mathrm{s2}}}{D_{\mathrm{s1}}\, D_{\mathrm{ds2}}}, \tag{2}$$

where $D_{\mathrm{ds1}}$, $D_{\mathrm{s2}}$, $D_{\mathrm{s1}}$ and $D_{\mathrm{ds2}}$ are the angular diameter distances of source 1 from the deflector, source 2 from the observer, source 1 from the observer, and source 2 from the deflector, respectively (see Figure 1). $\beta_{12}$ is often referred to as the ratio of Einstein radii between source 1 and source 2 (e.g., T. E. Collett & M. W. Auger 2014), as the ratio of Einstein radii squared is proportional to $\beta_{12}$ for a singular isothermal mass profile.

### 2.1. Constraining Cosmology

Each angular diameter distance in Equation (2) is a function of redshift $z$ and the cosmological parameters $H_0$, $\Omega_{\mathrm{m}}$, and $w$ under the assumption of a flat $w$CDM cosmology, as shown by:

$$D_{ij} = \frac{c}{H_0(1 + z_j)} \int_{z_i}^{z_j} \frac{dz}{E(z)}, \tag{3}$$

where the normalized Hubble parameter $E(z) \equiv H(z)/H_0$ is given by:

$$E(z) = \sqrt{\Omega_{\mathrm{m}}(1 + z)^3 + \Omega_\Lambda(1 + z)^{3(1+w)}}, \tag{4}$$

with $\Omega_{\mathrm{m}} + \Omega_\Lambda = 1$ under the assumption of a flat Universe (i.e., $\Omega_k = 0$). Due to the direct inverse proportionality between $D_{ij}$ and $H_0$, the ratio of distances in $\beta_{ij}$ is independent of $H_0$. Therefore, with an independent measurement of $\beta_{12}$, we can constrain $\Omega_{\mathrm{m}}$ and $w$ provided we have redshift measurements of each component galaxy. We discuss more details of how we sample $\beta_{12}$ in Section 4.

Equation (4) allows us to describe two possible cosmologies: flat $\Lambda$CDM, where we fix $w = -1$, enabling us to constrain $\Omega_{\mathrm{m}}$ alone, and flat $w$CDM, where $w$ remains a free parameter. Within $\Lambda$CDM, the cosmological constant $\Lambda$ denotes dark energy as a homogeneous and isotropic fluid with a constant energy density over time. The condition that the dark energy density remains constant with the expansion of the Universe requires $w = -1$. However, removing this assumption allows us to constrain more flexible models governed by a flat $w$CDM cosmology where $w$ is free to take any constant value within our adopted prior bounds (i.e., $w \in [-2, 0]$).

We choose to examine cosmological models with a constant dark energy equation of state parameter because of the small sample size of only three DSPLs. However, DSPL cosmography is not limited to this choice. Any model with an analytically defined $E(z)$ can be substituted into Equation (3). The $w_0w_a$CDM model (M. Chevallier & D. Polarski 2001; E. V. Linder 2003) is an example that uses a time-dependent parameterization of the dark energy equation of state. Although introducing additional parameters would grant the model more freedom, it would also increase uncertainties for each parameter and introduce degeneracies between them, as they are not well constrained by our current sample size.

### 2.2. Mass-sheet Degeneracy

The mass-sheet degeneracy (MSD) refers to the degeneracy in the mass model that manifests itself through the so-called mass-sheet transformation of the mass model, accompanied by an associated scaling of the source position, that leaves all imaging observables invariant (E. Falco et al. 1985; M. Gorenstein et al. 1988). MSD allows transformations in the mass of the deflectors without altering our model reconstruction, leading to changes of $\mathcal{O}(10^{-2})$ in the value of $\beta_{12}$ measured from lens models (P. Schneider 2014). We discuss how MSD impacts our lens model in Section 5.

## 3. Observations

AGEL0353, also known as DESJ0353−1706, was identified as a strong lens candidate by C. Jacobs et al. (2019a, 2019b) using convolutional neural networks on Dark Energy Survey (DES; DES Collaboration 2016) imaging. T. M. Barone et al. (2025) spectroscopically confirmed that AGEL0353 is a strong gravitational lens and identified a second source with a distinct redshift from the first, classifying it as a DSPL. They used high-resolution imaging from the Hubble Space Telescope (HST) to resolve the lensed arcs of the background sources, as shown in Figure 2. More information on the HST imaging and spectroscopic measurements is provided in Sections 3.2 and 3.3.

We select AGEL0353 from six DSPLs presented in T. M. Barone et al. (2025). We include another of these six, AGEL150745+052256 (hereafter AGEL1507), in our combined DSPL constraints presented in Section 6.2, using the lens model and cosmological inference presented by N. Sahu et al. (2025). In addition, we include the DSPL J0946 (i.e., the Jackpot lens; modeled by T. E. Collett & M. W. Auger 2014; R. J. Smith & T. E. Collett 2021) in our combined constraints.

### 3.1. System Configuration

AGEL0353 consists of four galaxies: the deflector (G1), source 1 (S1), source 2a (S2a), and source 2b (S2b). Each system component is identified in Figure 2. Upon first glance, the lensed image pairs from S1 and S2b could represent an Einstein cross from a single source. However, our spectroscopic observations are insufficient to confirm or reject this (see Section 3.3). During the lens modeling process, detailed in Section 4.2, we were unable to reproduce these four images using a single source. Instead, our model indicates that the pair of images, outlined by the dashed white lines, originates from an independent source (S2b) with an assumed redshift equal to that of S2a. Therefore, this system is divided into three redshift planes: the deflector plane at $z_{\rm defl} = 0.617$, the first source plane at $z_{S1} = 1.460$, and the second source plane at $z_{S2} = 1.675$.

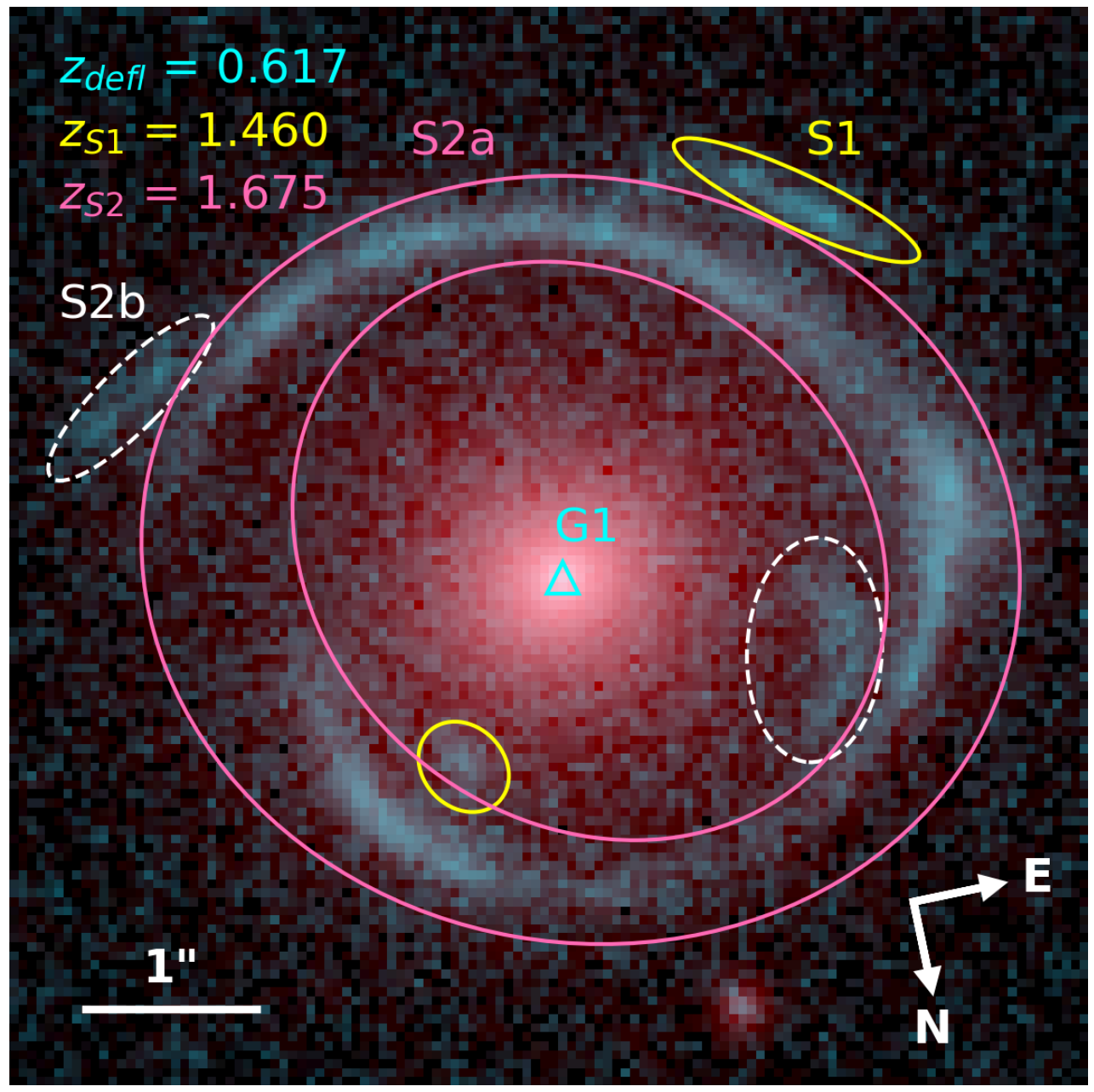

**Figure 2.** Color composite of AGEL0353 in the F200LP (blue) and F140W (red) filter bands from HST's Wide Field Camera 3 (WFC3) with sides of 6″. The cyan triangle marks the center of G1. The yellow lines encircle the pair of images corresponding to S1. The pink lines encircle the arc created by S2a. The dashed white lines outlined the pair of images corresponding to S2b. The measured redshifts for G1, S1, and S2a are annotated in corresponding colored texts. We assume the redshift of S2b to be equal to $z_{S2}$.

### 3.2. Imaging

AGEL0353 was imaged with the HST Wide Field Camera 3 (WFC3) in the F200LP (UVIS) and F140W (IR) filters, shown in Figure 2. Observations were made during the snapshot program SNAP–16773 (Cycle 29, PI: K. Glazebrook) using one 300 s exposure for the F200LP and three 300 s exposures for the F140W. The HST observations can be accessed from the Mikulski Archive for Space Telescopes (MAST) at the Space Telescope Science Institute via DOI: 10.17909/sb1r-er38. Data were reduced using the ASTRODRIZZLE task from the DRIZZLEPAC 2.0 software package (R. J. Avila et al. 2015), using 0.″05 and 0.″08 pixel scales for the F200LP and F140W images, respectively. We use the LACOSMIC[21] package, an implementation of the L.A. cosmic algorithm (P. G. van Dokkum 2001), to identify and remove cosmic rays from the F200LP observation.

We obtain each filter's point-spread function (PSF) using Tiny Tim (J. E. Krist et al. 2011). We select the F200LP observation for lens modeling because the lensed images of S1 and S2b are blended with the light of G1 in the F140W filter image, which may introduce bias into the lens model. The F200LP filter, which lies in the UV–visible range, introduces more complexity into the light profile of the sources over the F140W, which lies in the near-IR. This is because sources at $z \sim 1.5$ appear clumpier due to the presence of star-forming regions that emit strongly in the UV (C. J. Conselice 2014).

[21] https://lacosmic.readthedocs.io/en/stable/

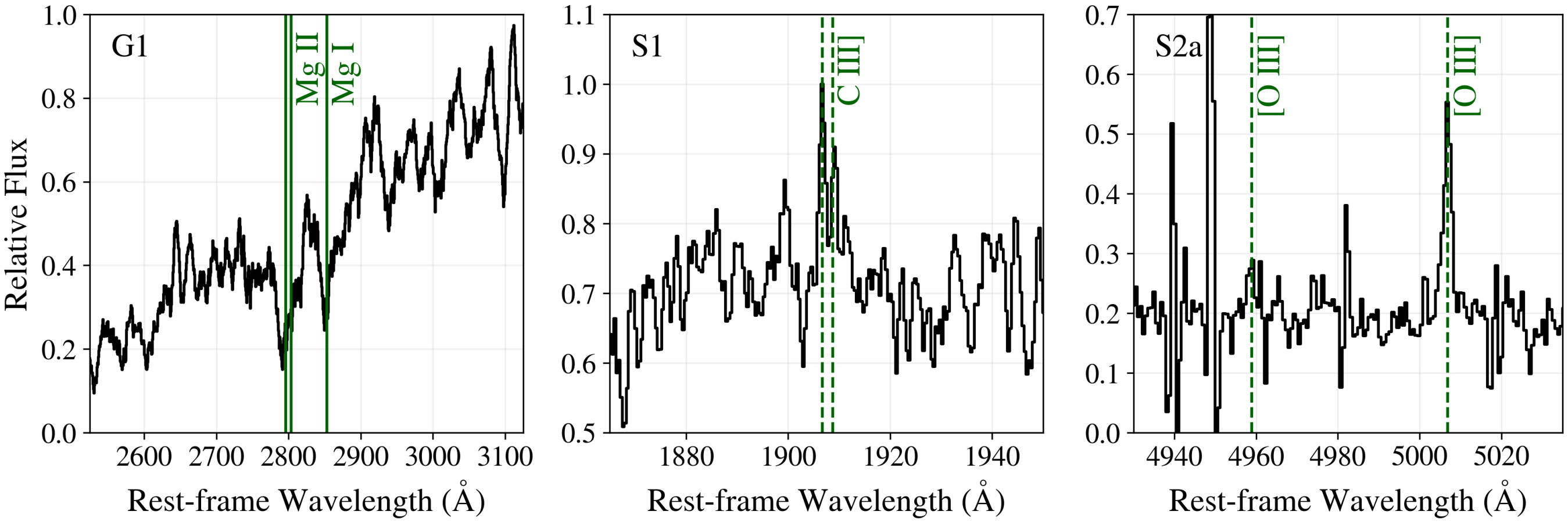

**Figure 3.** Rest-frame spectra for (from left to right) G1, S1 and S2a of AGEL0353. The integrated spectra of G1 and S1 used to measure redshifts of $z_{\rm defl} = 0.617$ and $z_{\rm S1} = 1.460$ were obtained using KCWI integral field spectroscopy. The spectrum of S2a used to measure a redshift of $z_{\rm S2} = 1.675$ was obtained using NIRES. The green vertical lines display the emission (dashed) and absorption (solid) lines used in determining each redshift.

### *3.3. Spectroscopy*

T. M. Barone et al. (2025) extracted the spectroscopic redshifts for G1 and S1 from the spectra (left and middle panels of Figure 3) obtained by the Keck Cosmic Web Imager (KCWI; P. Morrissey et al. 2018) during the 2021B U013 program (PI: T. Jones). These authors used integral field spectroscopic data (Figure 4) obtained from KCWI using the medium slicer and the blue low-resolution grating that spans the 3500–5600 Å wavelength range with a spectral resolution $R \approx 1800$. These authors used a $16\farcs5 \times 20\farcs4$ field of view with $0\farcs7 \times 0\farcs3$ spatial pixels and obtained a single exposure of 1200 s, and reduced the data using the same procedure described in K. V. GC et al. (2025).

The redshift for S2a was extracted by T. M. Barone et al. (2025) from the spectrum (right panel of Figure 3) obtained by Keck's Near-Infrared Echellette Spectrometer (NIRES; J. C. Wilson et al. 2004) during the 2021B W242 program (PI: G. Kacprzak). With NIRES, these authors used a $0\farcs55 \times 18''$ long slit with a $0\farcs15$ pixel scale, covering a 9000–24500 Å wavelength range with a spectral resolution $R = 2700$. To extract the 1D spectrum of S2a, they used the NIRES Spectral eXtraction (NSX) reduction pipeline.[22]

In the KCWI spectrum of G1, we see Mg I and Mg II absorption lines corresponding to $z_{\rm defl} = 0.617$. We see the C III] doublet emission in the KCWI spectrum of S1, which corresponds to $z_{\rm S1} = 1.460$. In the median KCWI image of AGEL0353, shown in Figure 4, the regions in which these lines were identified from integrated spectra are highlighted in cyan and yellow for G1 and S1, respectively. The brightest image of S1 and the arc of S2a cannot be resolved from one another by the IFU; hence, T. M. Barone et al. (2025) selected the yellow region to extract spectral features from both S1 and S2a. In the NIRES spectrum of S2a, we see [O III] emission lines at $z_{\rm S2} = 1.675$. The additional emission lines seen in the spectrum of S2a were identified as sky lines from the atmosphere. We do not have a spectroscopic redshift for S2b. As mentioned in Section 3.1, based on our lens modeling, we assume that S2b is at the same redshift as S2a.

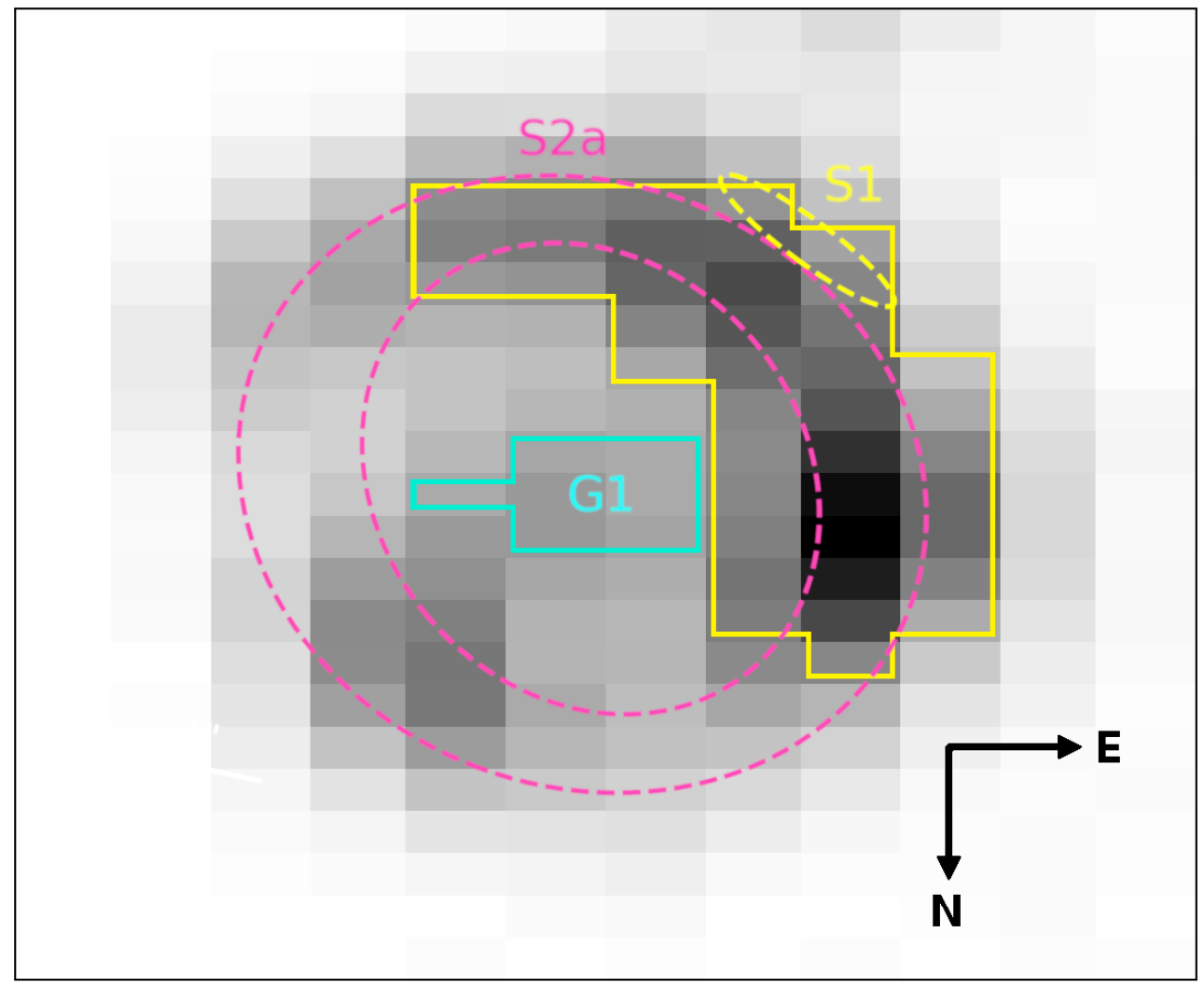

**Figure 4.** Median image of AGEL0353 from KCWI's medium slicer and blue low-resolution grating. The cutout is 12 × 23 spatial pixels corresponding to a field of view of $8\farcs4 \times 6\farcs9$ with a spatial pixel size of $0\farcs7 \times 0\farcs3$. The solid blue and yellow outlined regions indicate which pixels were integrated over to obtain the 1D spectra for G1 and S1 shown in Figure 3. The dashed pink and yellow outlined regions are overlaid to highlight the relative image positions of S2a and S1 from the HST imaging in Figure 2.

## 4. Lens Modeling

We use the multipurpose strong lensing software package LENSTRONOMY[23] (S. Birrer & A. Amara 2018; S. Birrer et al. 2021) to construct a model for the multiplane lensing system, following the same procedure as N. Sahu et al. (2025). The lens model for AGEL0353 comprises parameterized mass profiles of the deflector G1, intermediate source S1, and light profiles of G1, S1, S2a, and S2b, which we describe in the following section.

### *4.1. Model Profiles*

To describe the mass distribution of G1, we use an elliptical power law (EPL) profile (N. Tessore & R. B. Metcalf 2015). For the mass distribution of S1, we use a singular isothermal ellipse (SIE). The EPL and SIE are defined by their center $x_0$, $y_0$,[24] complex ellipticity moduli $e_1$, $e_2$, and Einstein radius $\theta_{\rm E}$ (representative of the deflection scale). The distinction between the EPL and SIE is that the EPL features an additional parameter $\gamma$ that describes the power law slope, which is fixed to $\gamma = 2$ for the SIE. We use an external shear profile, characterized by a magnitude $\gamma_{\rm ext}$ and an angular direction $\phi_{\rm ext}$, to account for the possible contribution of mass in the line-of-sight environment and any residual angular complexity in the mass distribution of G1 (A. Etherington et al. 2024; D. Johnson et al. 2025).

[22] https://sites.astro.caltech.edu/~tb/nsx/

[23] https://github.com/lenstronomy/lenstronomy

The light distributions of G1 and each source are described using elliptical Sérsic profiles (J. L. Sersic 1968). The elliptical Sérsic profile is defined by a half-light radius $R_{\rm half}$, a Sérsic index $n$, a center $x_0$, $y_0$, and complex ellipticity moduli $e_1$, $e_2$. Further details of each profile and its parameters can be found in the LENSTRONOMY documentation.[25]

### 4.2. Modeling Process

Here, we provide an overview of some aspects of our modeling process for AGEL0353. We obtain a fiducial model using a fixed flat ΛCDM cosmology with $H_0 = 70$ km s$^{-1}$ Mpc$^{-1}$, $\Omega_{\rm m} = 0.3$, and $\Omega_\Lambda = 0.7$. To obtain our fiducial model, we first perform image-position modeling, following the process described in N. Sahu et al. (2025, their Section 4.2), to acquire the best-fit model parameters from the lensed image positions alone. We follow this with extended-source modeling, which utilizes the pixel-level information in the F200LP imaging data, allowing us to compute an image reconstruction for comparison with the observed image. After we achieve an optimized model with our fixed fiducial cosmology, we free the choice of cosmology and constrain the cosmological distance ratio $\beta_{12}$ while further optimizing the extended-source modeling. This $\beta_{12}$ we measure then allows us to constrain $\Omega_{\rm m}$ for a flat ΛCDM cosmology and $\Omega_{\rm m}$ and $w$ for a flat $w$CDM cosmology, as we describe later in Section 6.

To quantify the efficacy of our model reconstruction, we use a reduced chi-squared statistic $\chi^2_{\rm red}$ we calculate through LENSTRONOMY (S. Birrer & A. Amara 2018; S. Birrer et al. 2021) as the square of the difference between the observed and the reconstructed images divided by the variance per pixel and the total number of pixels evaluated. We initially search for the best-fit model parameters using the particle swarm optimization (J. Kennedy & R. Eberhart 1995). We sample the parameter space around the best-fit values using the Markov chain Monte Carlo (MCMC) method with EMCEE (D. Foreman-Mackey et al. 2013). We use 204 parallel chains and perform as many steps as needed to reach a converged model. The chain is assumed to be converged when the median and standard deviation of the MCMC chains for each parameter are in equilibrium for at least 1000 steps. Once we reach convergence, we perform an additional 5000 steps to ensure a sufficiently large set of sampled points for extracting the posterior statistics of our model parameters.

During our modeling, we examine whether the lensed images of S1 and S2b originate from a single source component at $z_{\rm S1} = 1.460$. However, a single Sérsic profile is unable to reproduce the observed image configuration. Therefore, we test whether using two source components (each described by a single Sérsic) could solve this issue. Without a redshift measurement, we cannot predict whether S2b is closest to the first or second source plane before lens modeling. After testing both scenarios during modeling, a model that can reconstruct each image from all sources is only achieved by placing this source at $z_{\rm S2} = 1.675$, hence the label "S2b." This results in a $\chi^2_{\rm red}$ reduction of 0.039, equivalent to a 2.8$\sigma$ difference, and the inclusion or exclusion of S2b entirely does not alter the posterior distributions of the model parameters of interest. Consequently, we adopt this assumption.

## 5. Modeling Results

The best-fit model we obtain for AGEL0353, with $\chi^2_{\rm red} = 1.035$, is illustrated in Figure 5. The F200LP band image we use for our model is shown in the top left panel, the reconstructed model image in the top middle panel, and the normalized residual map in the top right panel. The projected source reconstructions are shown in the bottom left panel with the inner caustics of the deflector and S1 overlaid in green. The convergence and magnification maps are displayed in the bottom middle panel and bottom right panel, respectively. Our model accurately reproduces the lensed image positions and overall shapes (see top middle panel). The normalized residual map (top right panel) represents the difference between the reconstructed and observed image divided by the standard deviation. The residuals we see are typical among lens models used for cosmography (e.g., A. J. Shajib et al. 2020, 2022; D. M. Williams et al. 2025). Within the residual map, there is a minor level of correlated structures within the arcs of S1 and S2a. Possible causes and implications of these structures in the residual map are discussed in Section 7.

The posterior distributions of the key parameters from the lens mass model are shown in a corner plot in Figure 6, which displays the joint 2D distribution of each pair of these parameters with their 1$\sigma$ (16th and 84th percentiles) and 2$\sigma$ (2nd and 98th percentiles) levels shown by the dark and light shaded regions, respectively. The parameters include the effective projected Einstein radius of G1 for S2a and S2b $\theta_{\rm E,defl}$, the slope $\gamma$ of the elliptical power law mass profile of G1, the effective projected Einstein radius of S1 $\theta_{\rm E,S1}$, and the cosmological distance ratio $\beta_{12}$. The posteriors of the four mass profile parameters are approximately Gaussian. The posterior of $\beta_{12}$ does not appear Gaussian and instead ramps toward its maximum allowed prior value. A small positive correlation is seen in the 2D distribution of $\gamma_{\rm ext}$ and $\gamma$ as reported in other studies (e.g., A. J. Shajib et al. 2022; A. Etherington et al. 2023). Within the 2D distribution of $\beta_{12}$ and $\theta_{\rm E}$, defl, there appears to be a negative correlation. This degeneracy arises from the proportionality between $\theta_{\rm E,defl}$ and $\sqrt{1/\beta_{12}}$.

We summarize our findings and quote the median and 1$\sigma$ uncertainties of the key model parameters as follows:

1. The effective projected Einstein radius of G1 for S2a and S2b is $\theta_{\rm E,\,defl} = 1\overset{\prime\prime}{.}917^{+0.004}_{-0.003}$.
2. The power law density slope for the mass distribution of G1 is $\gamma = 2.494^{+0.029}_{-0.030}$. This value exceeds that of all lenses modeled in the AGEL survey so far (N. Sahu et al. 2024, 2025), but falls within the range of values $\gamma \in [1.72,\ 2.62]$ obtained by W. Sheu et al. (2024),

[24] Relative to the F200LP image center, at RAJ2000 = 58.44268000 deg and DECJ2000 = −17.11090000 deg, where positive $x_0$ direction points to East and positive $y_0$ direction to North.

[25] https://lenstronomy.readthedocs.io/

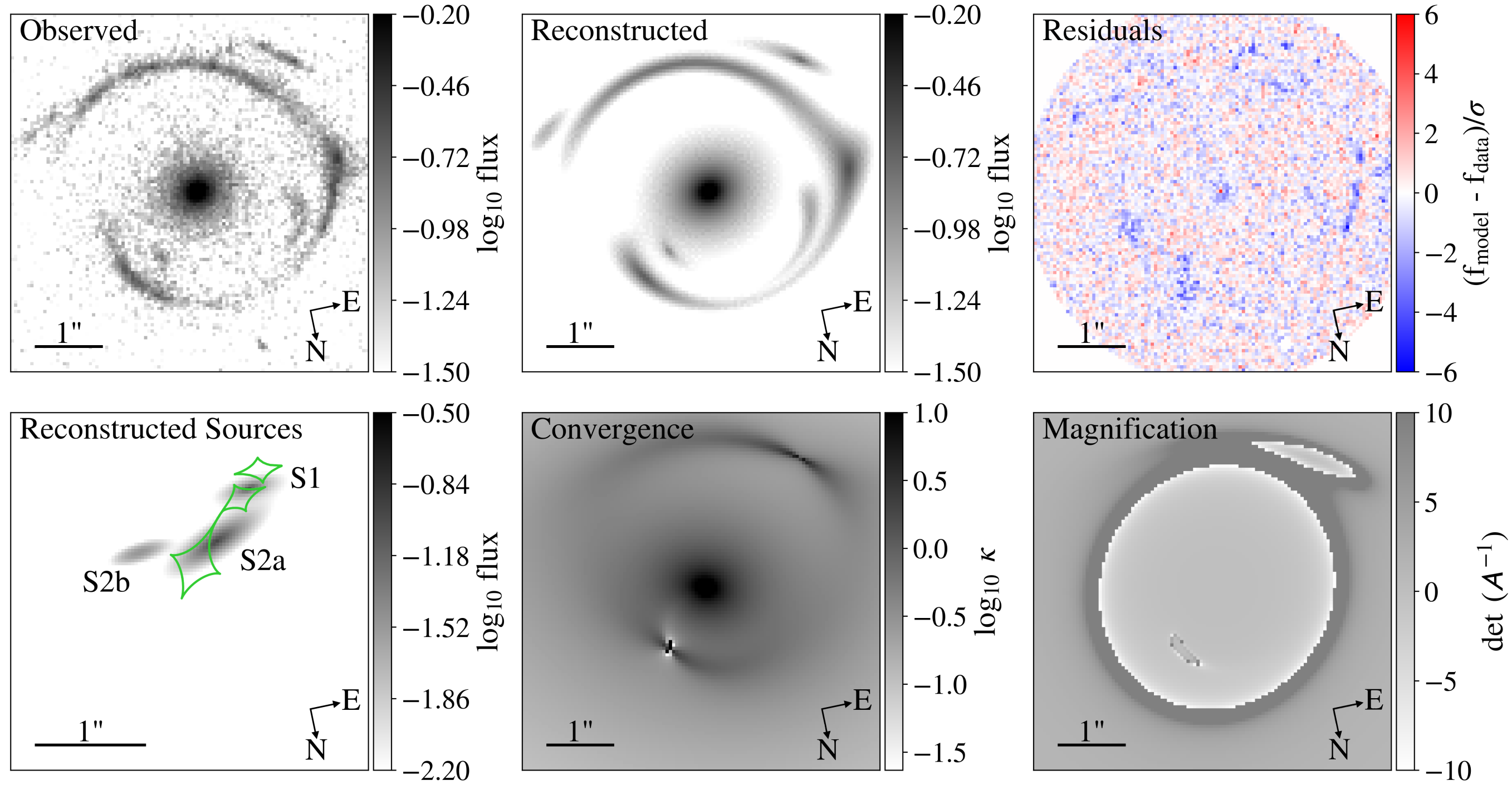

**Figure 5.** The best-fit model for AGEL0353. The observed image of AGEL0353 in HST's WFC3 F200LP filter is displayed in the top left panel, the reconstructed image produced by the lens model is in the top middle, and the normalized residual map between the observed and reconstructed images is displayed in the top right. The projected source reconstructions of S1, S2a, and S2b are shown in the bottom left panel with caustics (with respect to S2a's source plane) overlaid in green. The convergence map representing the projected lensing surface mass density of the system with respect to S2a is displayed in the bottom middle panel. The arc-like structures represent the projected convergence of S1, lensed by G1. The projected magnification with respect to S2a is shown in the bottom right panel. Similarly to the convergence, the arc-like structures represented the projected magnification of S1, lensed by G1. The model accurately reproduces the locations and overall shapes of the lensed images, which agree with their corresponding source-plane reconstructions.

C. Y. Tan et al. (2024) from lens models of elliptical galaxies at a mean redshift of 0.6 from the Strong Lensing Legacy Survey (SL2S; R. A. Cabanac et al. 2007; R. Gavazzi et al. 2012).

3. The magnitude of external shear $\gamma_{\rm ext} = 0.117^{+0.002}_{-0.002}$.
4. The effective projected Einstein radius of S1 is $\theta_{\rm E,\,s1} = 0\overset{''}{.}240^{+0.011}_{-0.016}$.
5. The cosmological distance ratio is $\beta_{12} = 0.929^{+0.004}_{-0.008}$, which lies within 1.9$\sigma$ of the fiducial value of 0.916.

The complete list of lens model parameters can be found in Appendix A (Table A1).

Physical properties of the deflector and sources in AGEL0353 are presented in Table 1, including total mass, apparent magnitude, and half-light radius. The point estimates are the medians calculated from our converged MCMC chains, and the stated 1$\sigma$ uncertainties are propagated from those of the lens model parameters. The total mass inferred from lens modeling is cosmology dependent; therefore, we adopt our inferred flat $w$CDM cosmology from the combination of three DSPLs and CMB constraints (see Section 6.3) for computing this quantity.

We test the impact of MSD on our posterior of $\beta_{12}$, which may be manifested through the presence of line-of-sight structures, by introducing a mass sheet to the G1 deflector plane. This consists of a single free parameter $\kappa_{\rm ext}$ that quantifies the external convergence of the mass sheet. We set the bounds $\kappa_{\rm ext} \in [-0.2, 0.2]$; a reasonable limit based on lensing + kinematic measurements of $\kappa_{\rm ext}$ that fall within this range (e.g., S. Birrer et al. 2025). We obtain the median and 1$\sigma$ uncertainties of $\kappa_{\rm ext} = -0.185^{+0.020}_{-0.010}$ and $\beta_{12} = 0.928^{+0.005}_{-0.008}$, closely consistent with the value of $\beta_{12} = 0.929^{+0.004}_{-0.008}$ obtained without a mass sheet. We find that MSD has a negligible impact on our cosmological inference; however, we do not address how a mass sheet in the plane of S1 would affect our result and leave this for further exploration in future studies. Therefore, we move forward with our model that does not include a mass sheet where $\beta_{12} = 0.929^{+0.004}_{-0.008}$.

## 6. Cosmological Constraints

Following Section 4.2, we obtain a posterior distribution of $\beta_{12}$ from the prior $\beta_{12} \in [0.911, 0.934]$; the possible range of values given that $w \in [-2, 0]$ and $\Omega_{\rm m} \in [0, 1]$. Using Equations (2) through Equation (4) and the same MCMC method used previously, we used our posterior distribution of $\beta_{12}$ and the spectroscopic redshifts of G1 ($z_{\rm defl}$), S1 ($z_{\rm S1}$), and S2a ($z_{\rm S2}$) to derive constraints on two adopted cosmologies: flat ΛCDM and flat $w$CDM. Using 200 parallel chains, we run the MCMC sampler until the chains converge for 4000 steps, and take our sample of $\Omega_{\rm m}$ and $w$ from the converged chains. Following Section 2.1, we constrain $\Omega_{\rm m}$ for a flat ΛCDM cosmology and constrain both $\Omega_{\rm m}$ and $w$ for a flat $w$CDM cosmology.

In the following sections, we first present our constraints on the flat ΛCDM and flat $w$CDM cosmologies from AGEL0353 alone, then combine our constraints with two other previously analyzed DSPLs, and external cosmological probes such as the CMB, SNe, and BAO. We combine independent constraints by approximating their posterior functions using a kernel density estimate (KDE), which are normalized so that they integrate to one (see Equation (10) in N. Sahu et al. 2025), and then multiply each KDE. Each median value of $\Omega_m$ and $w$ is presented with upper and lower limits that represent their $1\sigma$ uncertainties.

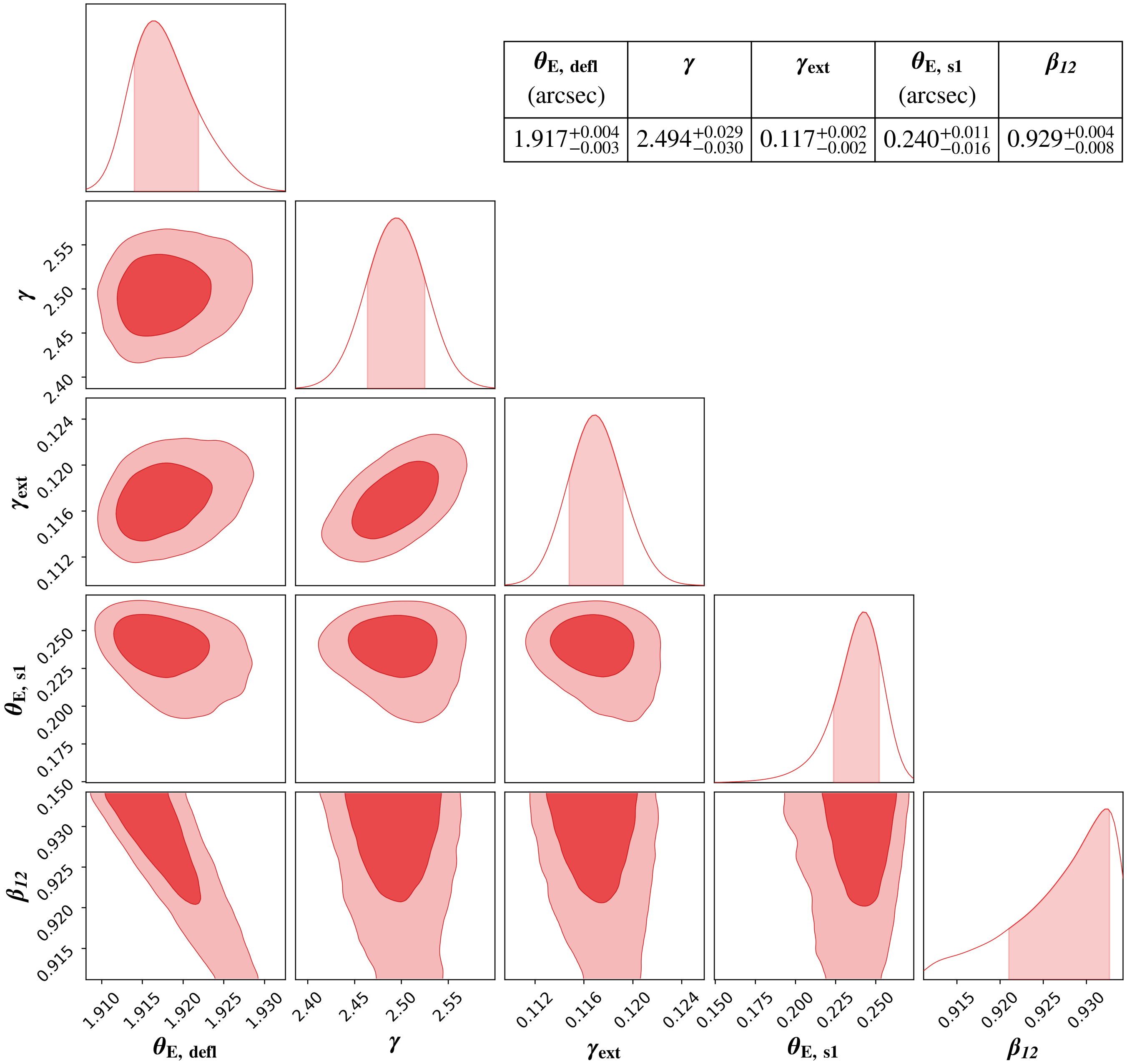

| $\theta_{E,\,defl}$ (arcsec) | $\gamma$ | $\gamma_{ext}$ | $\theta_{E,\,s1}$ (arcsec) | $\beta_{12}$ |
|---|---|---|---|---|
| $1.917^{+0.004}_{-0.003}$ | $2.494^{+0.029}_{-0.030}$ | $0.117^{+0.002}_{-0.002}$ | $0.240^{+0.011}_{-0.016}$ | $0.929^{+0.004}_{-0.008}$ |

**Figure 6.** Two-dimensional distributions for the Einstein radius $\theta_{E,defl}$ of G1, the power law density slope of G1 $\gamma$, the magnitude of external shear $\gamma_{ext}$, the Einstein radius of S1 $\theta_{E,s1}$ and the cosmological distance ratio $\beta_{12}$ from the final converged MCMC chains of our best-fit model for AGEL0353. The dark and light-shaded regions represent the 68% and 95% credible regions for each 2D distribution. In each panel along the diagonal from top left to bottom right, the red curve represents the posterior distribution of the corresponding parameter. The shaded region below each parameter's posterior distribution represents the $1\sigma$ credible region. Table (top right) lists the median and $1\sigma$ uncertainties for the posterior distributions of each parameter. The posterior distribution for each parameter appears approximately Gaussian except for $\beta_{12}$, which ramps toward its upper limit. There appear to be minor correlations between the $\beta_{12}$ and $\theta_{E,defl}$ and the $\gamma_{ext}$ and $\gamma$ distributions. The entire set of model parameters and their $1\sigma$ uncertainties can be found in Table A1.

### 6.1. Constraints from AGEL0353

Here, we present the posterior distribution of $\Omega_m$ for a flat $\Lambda$CDM cosmology using $\beta_{12}$ from our lens model of AGEL0353 (see Section 5). The probability density function (PDF) of $\Omega_m$ for flat $\Lambda$CDM is represented by the red curve in Figure 7, with the shaded region highlighting the 68% credible region. From the posterior, we compute a value of $\Omega_m = 0.265^{+0.324}_{-0.170}$.

Using the same method, we sample the posteriors of $\Omega_m$ and $w$ for a flat $w$CDM cosmology using $\beta_{12}$ from the AGEL0353 lens model. The joint distribution of $w$ and $\Omega_m$ is shown in the left panel of Figure 8. The constraints for $\Omega_m$ and $w$ are provided in Table 2.

### 6.2. Combined DSPL Constraints

We obtain combined DSPL constraints using three systems: AGEL0353 (this work), AGEL1507 from N. Sahu et al. (2025), and J0946 from T. E. Collett & M. W. Auger (2014).

**Table 1**
Spectroscopic Redshift, Total Mass, Apparent AB Magnitudes, and Half-light Radius $R_{\rm Half}$ of Each Component Galaxy in AGEL0353

| Component | SpectroscopicRedshift | Total Mass $\log_{10}$ $(M/M_\odot)$ | Magnitude (F200LP) | $R_{\rm half}$ (arcsec) |
|---|---|---|---|---|
| G1 | 0.617 | $12.27^{+0.03}_{-0.03}$ | $20.87^{+0.29}_{-0.24}$ | $2.56^{+0.32}_{-0.27}$ |
| S1 | 1.460 | $11.08^{+0.03}_{-0.02}$ | $25.44^{+0.46}_{-0.44}$ | $0.21^{+0.04}_{-0.03}$ |
| S2a | 1.675 | … | $24.84^{+0.11}_{-0.09}$ | $0.19^{+0.01}_{-0.01}$ |
| S2b | … | … | $26.54^{+0.15}_{-0.13}$ | $0.11^{+0.01}_{-0.01}$ |

**Note.** All spectroscopic redshifts have formal uncertainties of ±0.0005 (T. M. Barone et al. 2025), and the uncertainties of every other quantity are propagated from lens model parameters. Total mass is computed within three half-light radii for each component using the updated flat $w$CDM cosmology we find from our DSPL measurements combined with CMB constraints (see Section 6.3). Apparent magnitudes are computed from the F200LP data in the AB system and are corrected for magnification. $R_{\rm half}$ is the half-light radius parameter from the Sérsic profiles in each light model.

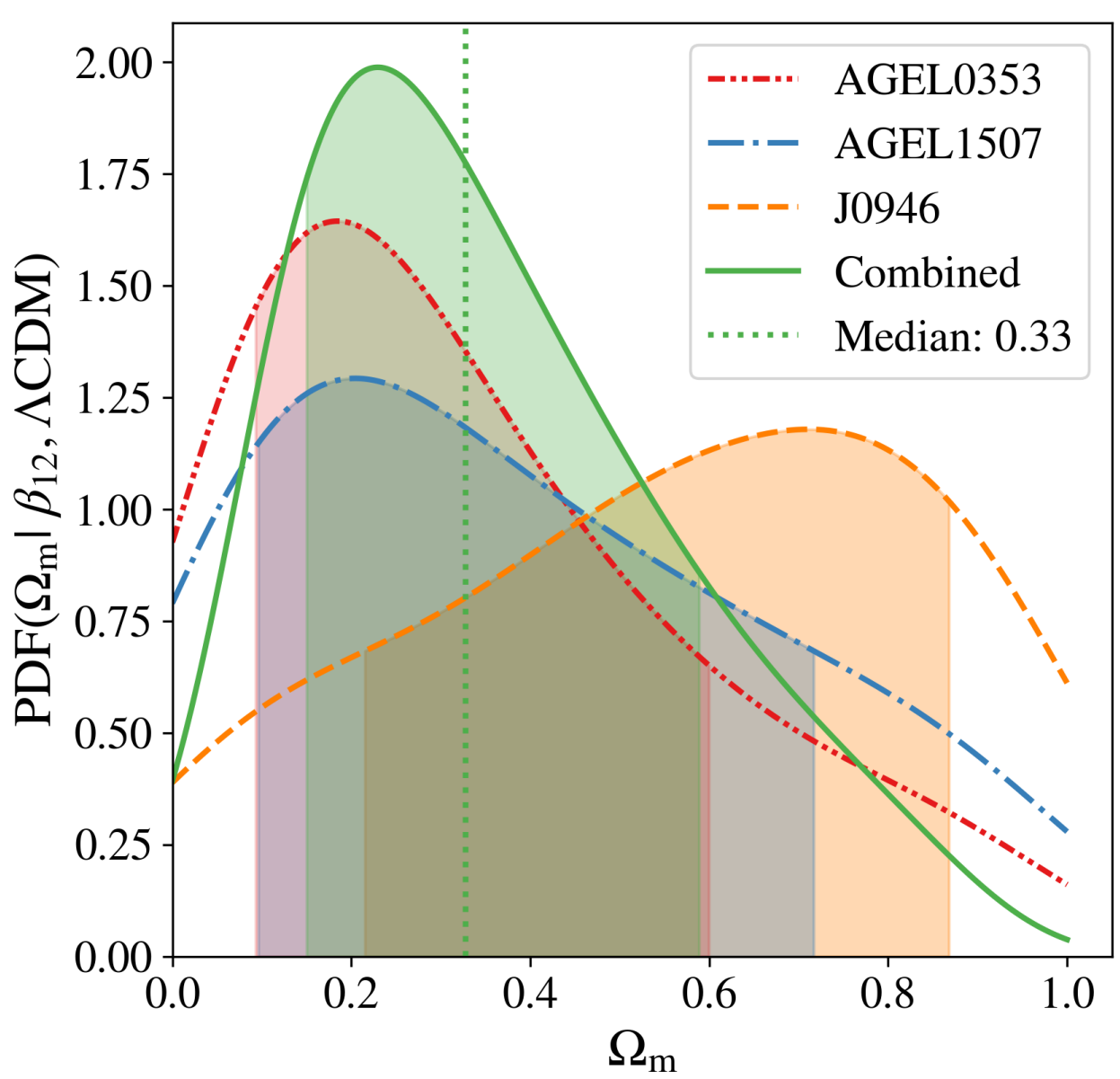

**Figure 7.** PDF of $\Omega_{\rm m}$ under the assumption of a flat ΛCDM cosmology given $\beta_{12}$ sampled from our lens model. The red, blue, and orange lines represent the individual PDFs for AGEL0353, AGEL1507, and J0946. The green line shows the combined PDF from all three DSPLs. The colored shaded regions represent the 68% credible interval for each PDF. The green dotted line indicates the median of the combined PDF.

For J0946, we use the updated constraints on $\Omega_{\rm m}$ and $w$ from N. Sahu et al. (2025, Section 6.3 therein) that use the updated spectroscopic redshift for "source 2" in J0946 from R. J. Smith & T. E. Collett (2021) and $\beta_{12}$ from T. E. Collett & M. W. Auger (2014).

For a flat ΛCDM cosmology, we combine the PDFs of $\Omega_{\rm m}$ from AGEL0353, AGEL1507, and J0946, as described previously. The PDFs for each DSPL and all three combined for the flat ΛCDM cosmology are shown in Figure 7. The constraints for $\Omega_{\rm m}$ are shown in Table 2 for each DSPL and the combined value. We observe a 12% improvement in the uncertainty of our combined value of $\Omega_{\rm m}$ over AGEL0353 alone.

For a flat $w$CDM cosmology, we present our combined DSPL constraints on $w$ and $\Omega_{\rm m}$ in the middle panel of Figure 8. The values of $\Omega_{\rm m}$ and $w$ for each DSPL and combined DSPL are presented in Table 2. From our combined measurement of three DSPLs, we see a 15% improvement in uncertainty for $w$ compared to AGEL0353 alone.

### *6.3. Constraints Combined with Other Cosmological Probes*

As we stated in Section 1, several standard probes are used to measure $\Omega_{\rm m}$ and $w$, including the CMB, Type Ia SNe, and BAO. By introducing a new independent probe, we can independently test existing constraints and provide tighter constraints by combining DSPL and existing constraints on $\Omega_{\rm m}$ and $w$. DSPLs are particularly suited to such a task because of the orthogonality between DSPL and CMB constraints (T. E. Collett et al. 2012; N. Sahu et al. 2025).

We combine the joint 2D constraints on $\Omega_{\rm m}$ and $w$ for the $w$CDM model from the sample of three DSPLs described above (Section 6.2) with the constraints from CMB measurements taken from the Planck Collaboration (2020). The two distributions from each probe and the combined distribution are shown in the right panel of Figure 8. The combined DSPL + CMB constraints for $\Omega_{\rm m}$ and $w$ are presented in Table 2. By combining DSPL and CMB constraints, we see an 11% and 39% improvement in uncertainty for $\Omega_{\rm m}$ and $w$, respectively, compared to CMB constraints alone.

For completeness, we take the Type Ia SNe constraints from the 5 yr DES SNe data set (DES Collaboration 2024) and the BAO constraints from the Dark Energy Spectroscopic Instrument (DESI) year 1 plus Sloan Digital Sky Survey (SDSS) observations (DESI Collaboration 2025a) and we compute the combined constraints from the DSPL, CMB, SNe, and BAO measurements to produce the tightest possible constraint, shown by the cyan region in Figure 9. The constraints for $\Omega_{\rm m}$ and $w$ for this combined constraint are presented in Table 2. This allows us to constrain $\Omega_{\rm m}$ and $w$ to within 3% uncertainty, assuming a flat $w$CDM cosmology. For now, the majority of constraining power comes from the standard CMB, SNe, and BAO measurements (as seen in Figure 9), which can achieve the same uncertainty in $w$ without DSPLs (DESI Collaboration 2025a). However, our final $w$ value after combining with DSPLs is higher by $\sim 1.2\sigma$ than that of DESI Collaboration (2025a), shifting our constraint on $w$ away from the ΛCDM model but remaining in agreement with it.

## 7. Discussion

### *7.1. Modeling Results*

The influence the choice of mass model has on the final constraints extracted from the model would benefit from a more detailed analysis using simulated data. For example, L. Van de Vyvere et al. (2022) explore the impact of

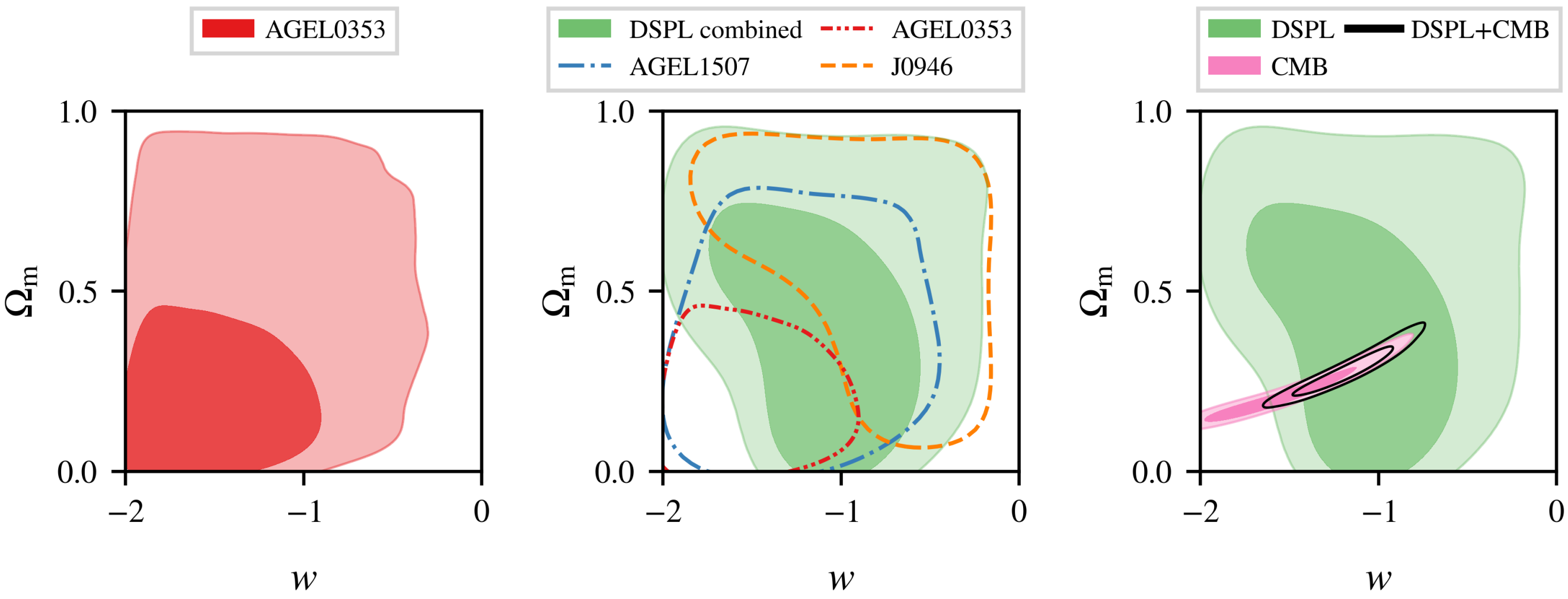

**Figure 8.** Two-dimensional distributions of $\Omega_m$ vs. $w$ assuming a flat $w$CDM cosmology. The dark and light-shaded regions represent the 68% and 95% credible regions for AGEL0353 alone, three DSPLs combined, and Planck CMB measurements (Planck Collaboration 2020) in red, green, and pink. The red, blue, and orange dashed contours represent the 68% credible levels for AGEL0353, AGEL1507, and J0946. The black contours represent the 68% and 95% credible levels for the combined constraints of the DSPLs and CMB measurements.

**Table 2**
Cosmological Parameter Constraints in the Case of a Flat ΛCDM and Flat $w$CDM Cosmologies from each DSPL, DSPLs Combined, DSPL, and CMB Measurements Combined, and DSPL, CMB, SNe, and BAO Measurements Combined

| **Flat ΛCDM** ($\Omega_k = 0$, $w = -1$) | | |
|---|---|---|
| | $\Omega_m$ | |
| AGEL0353 | $0.265^{+0.324}_{-0.170}$ | |
| AGEL1507 | $0.334^{+0.380}_{-0.234}$ | |
| J0946 | $0.597^{+0.269}_{-0.383}$ | |
| DSPLs (above) combined | $0.327^{+0.262}_{-0.178}$ | |
| Flat $w$CDM ($\Omega_k = 0$) | | |
| | $\Omega_m$ | $w$ |
| AGEL0353 | $0.192^{+0.305}_{-0.131}$ | $-1.52^{+0.49}_{-0.33}$ |
| AGEL1507 | $0.344^{+0.384}_{-0.239}$ | $-1.24^{+0.66}_{-0.51}$ |
| J0946 | $0.621^{+0.257}_{-0.386}$ | $-0.76^{+0.51}_{-0.99}$ |
| DSPLs (above) combined | $0.354^{+0.323}_{-0.253}$ | $-1.07^{+0.34}_{-0.36}$ |
| CMB | $0.198^{+0.068}_{-0.038}$ | $-1.57^{+0.35}_{-0.27}$ |
| DSPL + CMB | $0.275^{+0.050}_{-0.044}$ | $-1.18^{+0.19}_{-0.19}$ |
| DSPL + CMB + SNe + BAO | $0.316^{+0.008}_{-0.007}$ | $-0.96^{+0.03}_{-0.03}$ |

**Note.** The flat ΛCDM and $w$CDM constraints are shown in Figures 7 and 8, respectively. Figure 9 shows the combined constraints for all four independent observations (DSPL + CMB + SNe + BAO).

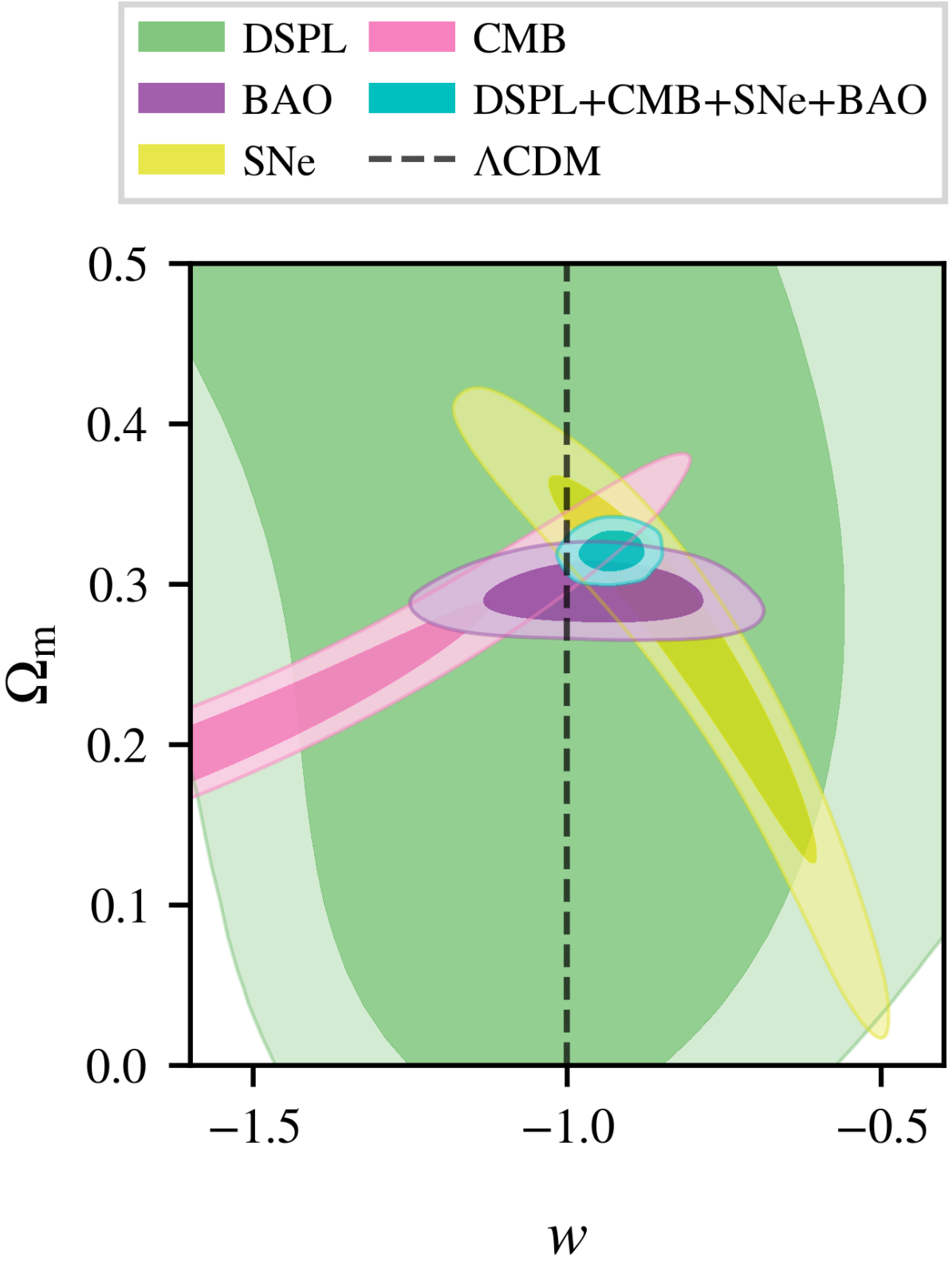

**Figure 9.** Two-dimensional distributions of $\Omega_m$ vs. $w$ under the assumption of a flat $w$CDM cosmology. The dark and light-shaded regions in green, purple, yellow, pink, and cyan represent the 68% and 95% credible regions for our DSPL, DESI year 1 BAO plus SDSS (DESI Collaboration 2025a), 5 yr DES SNe (DES Collaboration 2024), Planck CMB (Planck Collaboration 2020), and all the former measurements combined, respectively. The black dashed line represents values where $w = -1$ following a ΛCDM cosmology.

introducing a fourth-order multipole to account for disky or boxy structure within the mass profile and find that this does not significantly impact the value of $H_0$ they determine through time-delay cosmography. Similarly, for DSPLs, cosmology depends only on $\beta_{12}$, which should not be altered if the Einstein radius is well constrained. We measure $\theta_E$, defl to a precision of <1% and see no degeneracy between $\beta_{12}$ and $\gamma_{ext}$ (Figure 6), which could otherwise indicate the need for additional angular structure that may impact $\beta_{12}$. However, we do not test whether line-of-sight structures are in part responsible for the external shear we determine for our model.

D. Johnson et al. (2025) find that although line-of-sight structures have a subdominant effect on a large simulated sample of DSPL constraints, in the case of a flat $w_0w_a$CDM cosmology 35% of line-of-sight structure realizations introduced a $2\sigma$ bias to the cosmological parameters.

We quantify the impact that a mass sheet in the deflector plane has on our posterior of $\beta_{12}$ in Section 5 and find a change in uncertainty that can be attributed to random statistical variation, suggesting that it has a negligible impact. P. Schneider (2014) finds that although MSD can affect the value of $\beta_{12}$, its effect is less significant for DSPLs than for single-source systems because of the additional constraints on the deflector's mass profile from the second source. A simple way to mitigate this degeneracy is by using independently observed velocity dispersion of the deflectors (P. Schneider & D. Sluse 2013; N. Sahu et al. 2025). A caveat of our MSD investigation is that we do not address the impact of a mass sheet in the plane of S1; this would benefit from further investigation with the addition of velocity dispersion measurements of G1 and S1. However, we do not yet have the suitable spectroscopic data for AGEL0353 to make these velocity dispersion measurements.

The mass distributions of G1 and S1 allowed us to compute their total galaxy mass with precision $<1\%$ (see Table 1). However, the effect of MSD and mass along the line of sight makes the true uncertainty in the mass of S1 difficult to quantify. Stellar kinematic information, such as velocity dispersion measurements, would enable us to verify the masses of G1 and S1, and provide more robust mass estimates (S. Courteau et al. 2014). Therefore, there is likely some degeneracy between our computed masses and the choice of mass profile. However, assuming that an EPL and an SIE adequately describe the mass distributions of G1 and S1, they provide reasonable mass estimates.

### 7.2. Source Morphology

At first glance, the sets of lensed images produced by AGEL0353 appear to be the result of only two sources, with each producing a four-image configuration. However, we find that to reproduce these images, three sources are required when using an EPL plus external shear to describe the mass distribution of G1. The choice of redshift for S2b is explained in Section 4.2, where we only consider two possible redshift values during our modeling process. Alternatively, we could introduce the redshift of S2b as a free parameter. However, the separation between the lensed images of S2b is similar to that of S2a, indicating that they have comparable Einstein radii.

The structures seen in the residual map shown in Figure 5 are likely the result of structures within the light profile of the sources that are not captured by a smooth Sérsic profile. This is supported by the typical morphology of galaxies at $z \sim 1.5$, which have regions of star formation or spiral arms that cause perturbations in the light profile that are poorly described by a smooth Sérsic profile (C. J. Conselice 2014). Therefore, through the inclusion of a basis set of shapelets (S. Birrer et al. 2015) or pixelated source reconstruction, introducing perturbations to the Sérsic profile may be one way to remove these structures from the residuals.

### 7.3. Cosmological Constraints

In our fiducial cosmology, we assume that $H_0 = 70\ \mathrm{km\ s^{-1}\ Mpc^{-1}}$, which lies between the two values of $H_0$ determined from local and high-redshift observations (E. Di Valentino et al. 2021, 2025). A choice of $H_0$ is required to define the distances within our lens model. However, the choice of $H_0$ has no impact on our cosmological inference due to the independence of $\beta_{12}$, from which we constrain $\Omega_\mathrm{m}$ and $w$, on $H_0$ (see Section 2.1).

Our constraints from AGEL0353 are tighter for flat $\Lambda$CDM and $w$CDM cosmologies than AGEL1507 and J0946 (see Figure 7 and the middle panel of Figure 8). This is unexpected due to the less favorable source redshift configuration of AGEL0353 (T. E. Collett et al. 2012), compared to J0946 in particular (R. J. Smith & T. E. Collett 2021). However, this can be explained by the shape of the degeneracy between $\Omega_\mathrm{m}$ and $w$ constraints inferred from $\beta_{12}$ under a flat cosmology (see Appendix B). The value of $\beta_{12}$ that we measure is greater than that of J0946 relative to the fiducial values, resulting in some discrepancy in the inferred cosmology. It is unclear if this minor tension is due to systematic bias or some other effect.

### 7.4. Future Prospects

We increase the sample size of DSPLs from two to three, improving DSPL constraints on $\Omega_\mathrm{m}$ and $w$. Our sample of three DSPLs, when combined with other modern complementary probes such as the CMB, already provides a 39% improvement in uncertainty for $w$ over the CMB alone. This indicates significant potential for a larger sample of DSPLs, which could alone produce constraints comparable to those of the current standard probes of $w$CDM (D. Sharma et al. 2023; A. J. Shajib et al. 2024).

A larger DSPL sample also has the potential to explore alternative cosmological models with a time-evolving $w$, for example, the $w_0w_a$CDM model (M. Chevallier & D. Polarski 2001; E. V. Linder 2003) and the $w_\phi$CDM model (A. J. Shajib & J. A. Frieman 2025). To test whether DSPLs are viable probes of evolving dark energy models and $\Omega_k \neq 1$ scenarios, D. Sharma et al. (2023); A. J. Shajib et al. (2024) predicted constraints for future DSPL samples from the Rubin Observatory's Legacy Survey of Space and Time with size $\mathcal{O}(10^2)$. They compute a value of $w$ with an uncertainty of less than 27%. Once again, due to the orthogonality between DSPL and CMB constraints, their combination breaks the degeneracies between $w_0$, $w_a$, and other parameters.

In addition to the $\mathcal{O}(10^3)$ DSPL candidates Euclid is expected to identify throughout its wide field survey, $\mathcal{O}(10)$ even rarer triple-source-plane lenses (TSPLs) are predicted to be observed by Euclid (Euclid Collaboration 2025b). TSPLs could potentially be even more valuable cosmological probes due to their additional distance ratio parameters introduced by the third source. Efforts by D. J. Ballard et al. (2025, in preparation) are ongoing to obtain constraints from the only known galaxy-scale triple-source-plane system J0946, for which T. E. Collett & R. J. Smith (2020) discovered a third source.

A significant challenge for DSPL cosmography will be constructing robust lens models for the influx of newly discovered systems. We must ensure the robustness of models across all modeling methods and measure any biases they may introduce (e.g., A. J. Shajib et al. 2022; A. Galan et al. 2024). We must improve the efficiency of modeling DSPLs with the help of automation and machine learning methods (L. P. Levasseur et al. 2017; A. J. Shajib et al. 2025).

The AGEL survey's spectroscopic and HST imaging follow-up of 138 strong lenses has so far provided us with 6 DSPLs with complete redshifts (T. M. Barone et al. 2025). We aim to model the remaining AGEL DSPLs in future work to expand the sample of DSPLs used for cosmography. Our approach in building a high-quality strong lens sample is crucial for DSPL cosmography, which requires high spatial resolution imaging and multisource spectroscopy for each system. A key limiting factor in DSPL cosmography will be acquiring complete redshifts for each system, requiring multi-instrument spectroscopic follow-up on each DSPL. AGEL aims to tackle this problem with dedicated DSPL follow-ups to provide a benchmark sample of $\mathcal{O}(10)$ DSPLs for cosmography.

## 8. Conclusions

In this work, we analyze the DSPL AGEL0353 (Figure 2), inferring independent cosmological constraints from this system on flat $\Lambda$CDM and $w$CDM models (see Figures 7 and 8). We produce the first combined DSPL constraints from a sample of three systems using AGEL0353 and two previously analyzed systems, AGEL1507 (N. Sahu et al. 2025) and J0946 (T. E. Collett & M. W. Auger 2014; R. J. Smith & T. E. Collett 2021). We demonstrate the complementarity between the DSPL and CMB constraints, producing a 39% improvement in uncertainty when combining these two probes over the CMB data alone.

We construct a lens model of AGEL0353, shown in Figure 5, using the Python package LENSTRONOMY. We use an EPL and SIE to describe the mass profile of the deflector G1 and the intermediate source S1, respectively. Our lens modeling suggests that there are three sources, S1, S2a, and S2b; the latter two sources are in the farthest source plane (see Section 4). G1, S1 and S2a have measured spectroscopic redshifts of $z_{\rm defl} = 0.617$, $z_{\rm S1} = 1.460$, and $z_{\rm S2} = 1.675$. S2b is placed at an assumed redshift equal to $z_{\rm S2}$ (Section 3.3). From our lens modeling, we determined that the Einstein radius of G1 for S2a, and S2b is $\theta_{\rm E,defl} = 1.917^{+0.004}_{-0.003}$, with a power-law density slope $\gamma = 2.494^{+0.029}_{-0.030}$ for G1. Additionally, we find a lensing contribution from S1, which we determine to have an Einstein radius $\theta_{\rm E,s1} = 0.240^{+0.011}_{-0.016}$. The 2D distributions of the key model parameters are shown in Figure 6. Additional physical properties of the deflector and the sources, which we compute from our model, are summarized in Table 1.

Using $\beta_{12}$ from our lens model, we infer $\Omega_{\rm m} = 0.265^{+0.324}_{-0.170}$ for a flat $\Lambda$CDM model and $\Omega_{\rm m} = 0.192^{+0.305}_{-0.131}$ and $w = -1.52^{+0.49}_{-0.33}$ for a flat $w$CDM model from AGEL0353 alone. The combined constraints from AGEL0353, AGEL1507, and J0946 provide $\Omega_{\rm m} = 0.354^{+0.323}_{-0.253}$ and $w = -1.07^{+0.34}_{-0.36}$ for a flat $w$CDM model. We see a 15% improvement in $w$ compared to constraints from AGEL0353 alone.

We combine our DSPL constraints on $\Omega_{\rm m}$ and $w$ for a flat $w$CDM cosmology with CMB data, shown in Figure 8, providing $\Omega_{\rm m} = 0.275^{+0.050}_{-0.044}$ and $w = -1.18^{+0.19}_{-0.19}$. Combining the constraints of the sample of three DSPLS with CMB constraints delivers a 39% improvement in the uncertainty for $w$ compared to CMB constraints alone. We also present DSPL + CMB + SNe + BAO constraints for $w$CDM where $\Omega_{\rm m} = 0.316^{+0.008}_{-0.007}$ and $w = -0.96^{+0.03}_{-0.03}$. However, given the small sample of DSPLs so far, they provide little additional constraining power to the CMB + SNe + BAO constraints.

This paper demonstrates that DSPLs serve as an independent probe of cosmology, complementing other methods such as the CMB, motivating the expansion of the DSPL sample. Upcoming surveys from Euclid and Rubin will deliver $\mathcal{O}(10^3)$ DSPL candidates. At the same time, the follow-up spectroscopy from AGEL and 4SLSLS will verify hundreds of DSPL candidates usable for cosmography. This will enable us to improve the constraints on the models we have tested in this work and open the possibility of exploring more complex models to constrain the time evolution of the dark energy.

## Acknowledgments

This research was made possible by the Smithsonian Astrophysical Observatory (SAO)/University of Southampton Astronomy Masters Program. Parts of this research were conducted by the Australian Research Council Center of Excellence for All Sky Astrophysics in 3 Dimensions (ASTRO 3D), through project number CE170100013. Support for this work was provided by NASA through the NASA Hubble Fellowship grant HST-HF2-51492 awarded to AJS by the Space Telescope Science Institute (STScI), which is operated by the Association of Universities for Research in Astronomy, Inc., for NASA, under contract NAS5-26555. A.J.S. and H.S. were supported by NASA through the STScI grant HST-GO-16773, and A.J.S. also received support through the STScI grant JWST-GO-2974. T.M.B., K.V.G.C., D.J.B., and G.F.L. acknowledge support from the Australian Research Council Discovery Project grant DP230101775. T.J. and K.V.G.C. gratefully acknowledge financial support from NASA through grant HST-GO-16773, the Gordon and Betty Moore Foundation through Grant GBMF8549, the National Science Foundation through grant AST-2108515, and from a UC Davis Chancellor's Fellowship. KVGC was supported by NASA through the STScI grants JWST-GO-04265 and JWST-GO-03777. S.M.S. acknowledges funding from the Australian Research Council (DE220100003). S.H.S. thanks the Max Planck Society for support through the Max Planck Fellowship. This work has received funding from the European Research Council (ERC) under the European Union's Horizon 2020 research and innovation program (LensEra: grant agreement No. 945536). TEC is funded by the Royal Society through a University Research Fellowship. The authors thank the anonymous reviewer for their comments, which helped improve and ensure the quality of this paper.

*Facilities:* HST (WFC3), Keck:II (NIRES), Keck:II (KCWI).

*Software:* NUMPY (C. R. Harris et al. 2020), SCIPY (P. Virtanen et al. 2020), ASTROPY (Astropy Collaboration 2013, 2018, 2022), MATPLOTLIB (J. D. Hunter 2007), CHAINCONSUMER (S. Hinton 2016), GETDIST (A. Lewis 2019), MULTIPROCESS (M. M. McKerns et al. 2012).

## Appendix A
## AGEL0353 Model Parameters

Table A1 lists the median parameter values and their $1\sigma$ uncertainties we obtain from our lens model described in Section 5. A brief description of the model profiles and their

**Table A1**
Model Parameters for the Deflector and Sources of AGEL0353

| Components | Parameters | | | | | |
|---|---|---|---|---|---|---|
| Mass | | | | | | |
| EPL | $\theta_E$ (arcsec) | $\gamma$ | $e_1$ | $e_2$ | $x_0$ (arcsec) | $y_0$ (arcsec) |
| (G1) | $1.917^{+0.004}_{-0.003}$ | $2.494^{+0.029}_{-0.030}$ | $0.079^{+0.006}_{-0.006}$ | $0.068^{+0.006}_{-0.006}$ | $0.048^{+0.003}_{-0.003}$ | $-0.026^{+0.003}_{-0.003}$ |
| External shear | $\gamma_{ext}$ | $\varphi_{ext}$ | | | | |
| | $0.117^{+0.002}_{-0.002}$ | $0.500^{+0.012}_{-0.012}$ | | | | |
| SIE | $\theta_E$ (arcsec) | $e_1$ | $e_2$ | $x_0$ (arcsec) | $y_0$ (arcsec) | |
| (S1) | $0.240^{+0.011}_{-0.016}$ | $0.450^{+0.030}_{-0.035}$ | $-0.100^{+0.032}_{-0.031}$ | $0.777^{+0.013}_{-0.013}$ | $-0.859^{+0.016}_{-0.017}$ | |
| Lens Light | | | | | | |
| Elliptical Sérsic | $R_{half}$ (arcsec) | $n$ | $e_1$ | $e_2$ | $x_0$ (arcsec) | $y_0$ (arcsec) |
| (G1) | $2.560^{+0.318}_{-0.267}$ | $5.822^{+0.263}_{-0.245}$ | $0.071^{+0.006}_{-0.007}$ | $-0.050^{+0.007}_{-0.007}$ | $0.093^{+0.001}_{-0.001}$ | $0.014^{+0.001}_{-0.001}$ |
| Source Light | | | | | | |
| Elliptical Sérsic | $R_{half}$ (arcsec) | $n$ | $e_1$ | $e_2$ | $x_0$ (arcsec) | $y_0$ (arcsec) |
| (S1) | $0.213^{+0.041}_{-0.033}$ | $1.823^{+0.360}_{-0.303}$ | $0.450^{+0.030}_{-0.035}$ | $-0.100^{+0.032}_{-0.031}$ | $0.777^{+0.013}_{-0.013}$ | $-0.859^{+0.016}_{-0.017}$ |
| Elliptical Sérsic | $R_{half}$ (arcsec) | $n$ | $e_1$ | $e_2$ | $x_0$ (arcsec) | $y_0$ (arcsec) |
| (S2a) | $0.188^{+0.008}_{-0.007}$ | $0.981^{+0.052}_{-0.048}$ | $0.389^{+0.009}_{-0.009}$ | $-0.341^{+0.014}_{-0.013}$ | $0.377^{+0.009}_{-0.009}$ | $-0.417^{+0.014}_{-0.011}$ |
| Elliptical Sérsic | $R_{half}$ (arcsec) | $n$ | $e_1$ | $e_2$ | $x_0$ (arcsec) | $y_0$ (arcsec) |
| (S2b) | $0.113^{+0.007}_{-0.006}$ | $0.549^{+0.070}_{-0.035}$ | $0.496^{+0.003}_{-0.006}$ | $-0.113^{+0.027}_{-0.028}$ | $-0.341^{+0.016}_{-0.018}$ | $-0.466^{+0.011}_{-0.011}$ |

**Note.** Details of how each parameter value is extracted can be found in Section 4. The parameters $e_1$, $e_2$, $x_0$ and $y_0$ are joined between the mass (SIE) and light (Sérsic) profiles of S1, hence their values in each are identical.

parameters can be found in Section 4.1, and for more details of the chosen profiles, see the LENSTRONOMY documentation.[27]

## Appendix B
## Inferred Cosmological Uncertainties

For any given DSPL, the constraining power on $\beta_{12}$ is dependent upon its redshift configuration (T. E. Collett et al. 2012). AGEL0353 has a less favorable lensing configuration compared to J0946. Despite this, the uncertainties we infer from $\Omega_m$ and $w$ from AGEL0353 are smaller than those inferred from J0946 (Table 2). This change in uncertainty arises from the shape of the degeneracy between $\Omega_m$ and $w$.

Figure B1 compares two values of $\beta_{12}/\beta_{12,\mathrm{fid}}$ with equal percentage uncertainties for AGEL0353 and J0946, respectively. These values and their uncertainty are chosen to illustrate the change in the shape of the $\Omega_m - w$ degeneracy and are not physically significant. In each panel, the (red) blue region is (smaller) larger; thus, the inferred uncertainties on $\Omega_m$ and $w$ will be (smaller) larger for (higher) lower values of $\beta_{12}/\beta_{12,\mathrm{fid}}$ for a given lens. Therefore, the same percentage uncertainty in $\beta_{12}/\beta_{12,\mathrm{fid}}$ does not produce equally sized regions across the $\Omega_m - w$ parameter space.

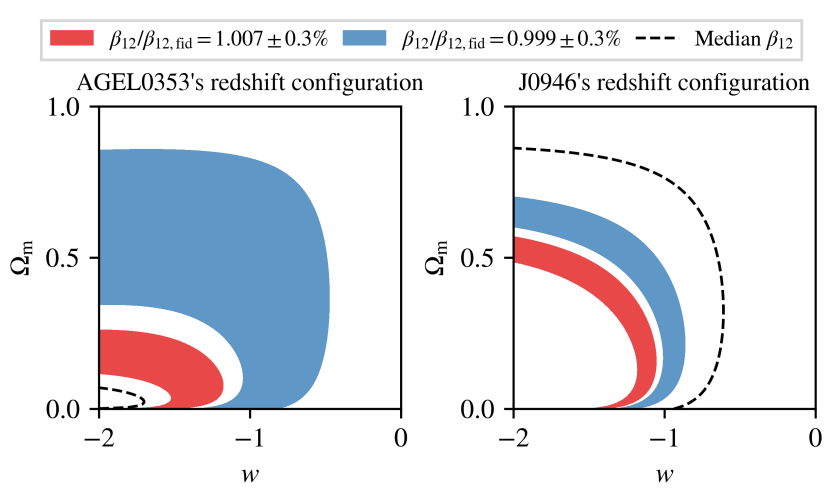

**Figure B1** Regions within the $\Omega_m - w$ parameter space for a flat $w$CDM cosmology. In the left and right panels, the possible range of $\beta_{12}$ is given for the lensing configurations of AGEL0353 and J0946, where $\beta_{12} \in [0.911, 0.934]$ and $\beta_{12} \in [0.706, 0.763]$, respectively. AGEL0353 and J0946 each have distinct values of $\beta_{12,\mathrm{fid}} = 0.916$ and $\beta_{12,\mathrm{fid}} = 0.721$, respectively. Red and blue regions are bounded by an uncertainty of $\pm 0.3\%$ on two values of $\beta_{12}/\beta_{12,\mathrm{fid}}$. In red are the regions centered on $\beta_{12}/\beta_{12,\mathrm{fid}} = 1.007$ and in blue are the regions centered on $\beta_{12}/\beta_{12,\mathrm{fid}} = 0.999$. The black dashed lines represent where the median values of $\beta_{12}$, which we measure from lens models of AGEL0353 and J0946, exist within the parameter space.

## ORCID iDs

Duncan J. Bowden https://orcid.org/0009-0008-6114-1401
Nandini Sahu https://orcid.org/0000-0003-0234-6585
Anowar J. Shajib https://orcid.org/0000-0002-5558-888X
Kim-Vy Tran https://orcid.org/0000-0001-9208-2143
Tania M. Barone https://orcid.org/0000-0002-2784-564X
Keerthi Vasan G. C. https://orcid.org/0000-0002-2645-679X
Daniel J. Ballard https://orcid.org/0009-0003-3198-7151
Thomas E. Collett https://orcid.org/0000-0001-5564-3140
Faith Dalessandro https://orcid.org/0009-0000-4328-200X
Giovanni Ferrami https://orcid.org/0000-0002-2012-4612
Karl Glazebrook https://orcid.org/0000-0002-3254-9044
William J. Gottemoller https://orcid.org/0009-0008-5372-1318
Leena Iwamoto https://orcid.org/0009-0006-4812-2033
Tucker Jones https://orcid.org/0000-0001-5860-3419
Glenn G. Kacprzak https://orcid.org/0000-0003-1362-9302
Geraint F. Lewis https://orcid.org/0000-0003-3081-9319
Haven McIntosh-Lombardo https://orcid.org/0009-0000-9974-5683
Hannah Skobe https://orcid.org/0000-0003-0516-3485
Sherry H. Suyu https://orcid.org/0000-0001-5568-6052

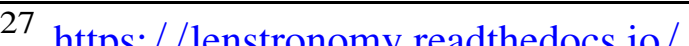

[27] https://lenstronomy.readthedocs.io/

Sarah M. Sweet https://orcid.org/0000-0002-1576-2505

## References

Astropy Collaboration 2013, A&A, 558, A33
Astropy Collaboration 2018, AJ, 156, 123
Astropy Collaboration 2022, ApJ, 935, 167
Avila, R. J., Hack, W., Cara, M., et al. 2015, in ASP Conf. Ser. 495, Astronomical Data Analysis Software and Systems XXIV, ed. A. R. Taylor & E. Rosolowsky (San Francisco, CA: ASP), 281
Barone, T. M., Tran, K.-V., Kacprzak, G. G., et al. 2025, arXiv:2503.08041
Birrer, S., & Amara, A. 2018, PDU, 22, 189
Birrer, S., Amara, A., & Refregier, A. 2015, ApJ, 813, 102
Birrer, S., Buckley-Geer, E. J., Cappellari, M., et al. 2025, arXiv:2506.03023
Birrer, S., Shajib, A. J., Gilman, D., et al. 2021, JOSS, 6, 3283
Cabanac, R. A., Alard, C., Dantel-Fort, M., et al. 2007, A&A, 461, 813
Caminha, G., Rosati, P., Grillo, C., et al. 2019, A&A, 632, A36
Caminha, G., Suyu, S., Grillo, C., & Rosati, P. 2022, A&A, 657, A83
Chevallier, M., & Polarski, D. 2001, IJMPD, 10, 213
Collett, T. E. 2015, ApJ, 811, 20
Collett, T. E., & Auger, M. W. 2014, MNRAS, 443, 969
Collett, T. E., Auger, M. W., Belokurov, V., Marshall, P. J., & Hall, A. C. 2012, MNRAS, 424, 2864
Collett, T. E., & Smith, R. J. 2020, MNRAS, 497, 1654
Collett, T. E., Sonnenfeld, A., Frohmaier, C., et al. 2023, Msngr, 190, 49
Conselice, C. J. 2014, ARA&A, 52, 291
Courteau, S., Cappellari, M., de Jong, R. S., et al. 2014, RvMP, 86, 47
DES Collaboration 2016, MNRAS, 460, 1270
DES Collaboration 2022, PhRvD, 105, 023520
DES Collaboration 2024, ApJL, 973, L14
DESI Collaboration 2025a, JCAP, 2025, 021
DESI Collaboration 2025b, PhRvD, 112, 083515
Di Valentino, E., Levi Said, J., Riess, A., et al. 2025, PDU, 49, 101965
Di Valentino, E., Mena, O., Pan, S., et al. 2021, CQGra, 38, 153001
Etherington, A., Nightingale, J. W., Massey, R., et al. 2023, MNRAS, 521, 6005
Etherington, A., Nightingale, J. W., Massey, R., et al. 2024, MNRAS, 531, 3684
Euclid Collaboration 2025a, arXiv:2503.15302
Euclid Collaboration 2025b, arXiv:2503.15327
Falco, E., Gorenstein, M., & Shapiro, I. 1985, ApJL, 289, L1
Ferrami, G., & Wyithe, S. 2024, MNRAS, 532, 1832
Foreman-Mackey, D., Hogg, D. W., Lang, D., & Goodman, J. 2013, PASP, 125, 306
Galan, A., Vernardos, G., Minor, Q., et al. 2024, A&A, 692, A87
Gavazzi, R., Treu, T., Marshall, P. J., Brault, F., & Ruff, A. 2012, ApJ, 761, 170
GC, K. V., Jones, T., Shajib, A. J., et al. 2025, ApJ, 981, 105
Gorenstein, M., Falco, E., & Shapiro, I. 1988, AJ, 327, 693
Harris, C. R., Millman, K. J., van der Walt, S. J., et al. 2020, Natur, 585, 357
Hinton, S. 2016, JOSS, 1, 45
Hunter, J. D. 2007, CSE, 9, 90
Jacobs, C., Collett, T., Glazebrook, K., et al. 2019a, ApJS, 243, 17
Jacobs, C., Collett, T., Glazebrook, K., et al. 2019b, MNRAS, 484, 5330
Johnson, D., Collett, T., Li, T., & Fleury, P. 2025, JCAP, 2025, 067
Jullo, E., Natarajan, P., Kneib, J.-P., et al. 2010, Sci, 329, 924
Kennedy, J., & Eberhart, R. 1995, in Proc. ICNN'95—Int. Conf. on Neural Networks, 4 (Piscataway, NJ: IEEE), 1942
Krist, J. E., Hook, R. N., & Stoehr, F. 2011, Proc. SPIE, 8127, 166
Levasseur, L. P., Hezaveh, Y. D., & Wechsler, R. H. 2017, ApJL, 850, L7
Lewis, A. 2019, JCAP, 2025, 025
Li, T., Collett, T. E., Krawczyk, C. M., & Enzi, W. 2024, MNRAS, 527, 5311
Linder, E. V. 2003, PhRvL, 90, 091301
Mahler, G., Jauzac, M., Richard, J., et al. 2023, ApJ, 945, 49
McKerns, M. M., Strand, L., Sullivan, T., Fang, A., & Aivazis, M. A. G. 2012, arXiv:1202.1056
Morrissey, P., Matuszewski, M., Martin, D. C., et al. 2018, ApJ, 864, 93
Perivolaropoulos, L., & Skara, F. 2022, NewAR, 95, 101659
Planck Collaboration 2020, A&A, 641, A6
Roszkowski, L., Sessolo, E. M., & Trojanowski, S. 2018, RPPh, 81, 066201
Sahu, N., Shajib, A. J., Tran, K.-V., et al. 2025, ApJ, 991, 72
Sahu, N., Tran, K.-V., Suyu, S. H., et al. 2024, ApJ, 970, 86
Schneider, P. 2014, A&A, 568, L2
Schneider, P., Ehlers, J., & Falco, E. E. 1992, Gravitational Lenses, Astronomy and Astrophysics Library (Berlin: Springer)
Schneider, P., Kochanek, C. S., & Wambsganss, J. 2006, Gravitational Lensing: Strong, Weak and Micro (Berlin: Springer)
Schneider, P., & Sluse, D. 2013, A&A, 559, A37
Scolnic, D. M., Jones, D. O., Rest, A., et al. 2018, ApJ, 859, 101
Sersic, J. L. 1968, Atlas de Galaxias Australes (Cordoba: Observatorio Astronomico)
Shajib, A. J., Birrer, S., Treu, T., et al. 2020, MNRAS, 494, 6072
Shajib, A. J., & Frieman, J. A. 2025, PhRvD, 112, 063508
Shajib, A. J., Nihal, N. S., Tan, C. Y., et al. 2025, ApJ, 992, 40
Shajib, A. J., Smith, G. P., Birrer, S., et al. 2024, RSPTA, 383, 2295
Shajib, A. J., Wong, K. C., Birrer, S., et al. 2022, A&A, 667, A123
Sharma, D., Collett, T. E., & Linder, E. V. 2023, JCAP, 2023, 001
Sheu, W., Shajib, A. J., Treu, T., et al. 2024, MNRAS, 541, 1
Smith, R. J., & Collett, T. E. 2021, MNRAS, 505, 2136
Tan, C. Y., Shajib, A. J., Birrer, S., et al. 2024, MNRAS, 530, 1474
Tessore, N., & Metcalf, R. B. 2015, A&A, 580, A79
Tran, K.-V. H., Harshan, A., Glazebrook, K., et al. 2022, AJ, 164, 148
Van de Vyvere, L., Gomer, M. R., Sluse, D., et al. 2022, A&A, 659, A127
van Dokkum, P. G. 2001, PASP, 113, 1420
Vegetti, S., Koopmans, L., Auger, M., Treu, T., & Bolton, A. 2014, MNRAS, 442, 2017
Virtanen, P., Gommers, R., Oliphant, T. E., et al. 2020, NatMe, 17, 261
Williams, D. M., Treu, T., Birrer, S., et al. 2025, arXiv:2503.00099
Wilson, J. C., Henderson, C. P., Herter, T. L., et al. 2004, Proc. SPIE, 5492, 1295